\documentclass[preprint]{aastex}

\received{}
\revised{}
\accepted{}
\ccc{}
\cpright{}{}

\slugcomment{2002, AJ in press}
\shorttitle{Abundance Ratios of M71}
\shortauthors{Ram\'{\i}rez \etal}

\newcommand{\kms}{km~s$^{-1}$}

\newcommand{\etal}{{\it et al.\/}}
\newcommand{\teff}{$T_{eff}$}
\newcommand{\grav}{log($g$)}
\newcommand{\mtv}{$\xi$}
\newcommand{\ew}{$W_{\lambda}$}
\newcommand{\fe}{[Fe/H]}

\begin{document}

\title{Abundances in Stars from the Red Giant Branch Tip to Near the 
Main Sequence Turn Off in M71: III. Abundance Ratios
\altaffilmark{1}}

\author{Solange V. Ram\'{\i}rez \altaffilmark{2} and  
Judith G. Cohen\altaffilmark{2}.}

\altaffiltext{1}{Based on observations obtained at the
W.M. Keck Observatory, which is operated jointly by the California 
Institute of Technology, the University of California, and the
National Aeronautics and Space Administration.}

\altaffiltext{2}{Palomar Observatory, Mail Stop 105-24,
California Institute of Technology.}

\begin{abstract}

We present abundance ratios for 23 elements with respect to Fe
in a sample of stars with a wide range in luminosity from 
luminous giants to stars near the turnoff in a globular cluster.
Our sample of 25 stars in M71 includes 10 giant stars more luminous than the
red horizontal branch (RHB), 3 HB stars, 9 giant stars less luminous than 
the RHB, and 3 stars near the turnoff.
The analyzed spectra, obtained with HIRES at the Keck Observatory,
are of high dispersion (R=$\lambda / \Delta \lambda$=35,000).
We find that the neutron capture, the iron peak and the 
$\alpha-$element abundance ratios 
show no trend with \teff, and low scatter around the mean between the
top of the RGB and near the main sequence turnoff.
The $\alpha-$elements Mg, Ca, Si and Ti are overabundant relative to Fe.
The anti-correlation between O and Na abundances, 
observed in other metal poor globular clusters, is detected in our sample
and extends to the main sequence.
A statistically significant correlation between Al and Na abundances
is observed among the M71 stars in our sample, extending to $M_V = +1.8$,
fainter than the luminosity of the RGB bump in M5.
Lithium is varying, as expected, and Zr may be varying from star to star
as well.

M71 appears to have abundance ratios very similar to M5
whose bright giants were studied by \citet{iva01}, but seems to have 
a smaller amplitude of star-to-star variations at a given luminosity,
as might be expected from its higher metallicity.
Both extremely O poor, Na rich stars and extremely O rich, Na poor stars
such as are observed in M5 and in M13 are not present in our 
sample of M71 stars.

The results of our abundance analysis of 25 stars in M71 provide
sufficient evidence of abundance variations at unexpectedly low luminosities
to rule out the mixing scenario.  Either alone or, even more
powerfully, combined with other recent studies of C and N
abundances in M71 stars, the existence of such  abundance variations
cannot be reproduced within the context of our current understanding
of stellar evolution.

\end{abstract}


\section{INTRODUCTION}

Abundance determinations of stars in Galactic globular clusters can provide 
valuable information about important astrophysical processes such as
stellar evolution, stellar structure, Galactic chemical evolution and
the formation of the Milky Way. Surface stellar abundances of C, N, O,
and often Na, Mg, and Al are found to be variable among red giants within 
a globular cluster. 
The physical process responsible for these star-to-star element variations 
is still uncertain (see the reviews of Kraft 1994 and
Pinsonneault 1997, as well as Cohen \etal\ 2001, Paper I).  

In order to study the origin of the star-to-star abundance variations,
we have started a program to determine chemical abundances of the nearer
galactic globular cluster stars. 
\citet{coh01} presents the sample of stars in M71, the nearest globular 
cluster reachable from the northern hemisphere, and the atmosphere parameters
of the program stars.
Our sample includes stars over a large range in luminosity:
19 giant stars, 3 horizontal branch (HB) stars, and 3 stars near the main 
sequence
turnoff, in order to study in a consistent manner red giants, horizontal
branch stars, and stars at the main sequence turnoff.
Our second paper \citep[][Paper II]{ram01} discusses the iron abundance in M71.
We found that the \fe\ abundances from both Fe I (\fe\ = $-0.71 \pm 0.08$)
and Fe II (\fe\ = $-0.84 \pm 0.12$) lines agree with each other and with 
earlier determinations \citep{coh83,gra86,lee87,sne94}.
We also found that the \fe\ obtained from Fe I and Fe II lines is constant 
within the rather small uncertainties for this group of stars over the full 
range in effective temperature (\teff) and luminosity. 
In this third paper of this series, we present our results for abundances of 
23 atomic species in our sample of M71 stars.

\section{ATOMIC LINE PARAMETERS}

The abundance analysis is done using a current version of the LTE 
spectral synthesis program MOOG \citep{sne73}. 
A line list specifying the wavelengths, excitation
potentials, $gf$ values, damping constants, and equivalent widths for the 
observed lines is required. 
The provenance of the equivalent widths, $gf$ values and damping constants 
is discussed below. 

In addition, a model atmosphere for the \teff\
and surface gravity (\grav) appropriate for each star and a value for the
microturbulent velocity (\mtv) are also required for the abundance analysis.
We use the grid of model atmospheres from \citet{kur93a} with a metallicity of
\fe\ = $-$0.5 dex, based on earlier high dispersion iron abundance analysis of
M71 \citep[][Paper II]{coh83,gra86,lee87,sne94}.
\teff\ and \grav\ are derived from the 
broad-band photometry of the stars as described in 
Paper I. The photometric \teff\ has an 
error of $\pm$75 K for giants and $\pm$150 K for
dwarfs and \grav\ has an error of $\pm$0.2 dex.
The microturbulent velocity is derived as described in Paper II; \mtv\ has
an error of $\pm$0.2 \kms. Table~\ref{tab1}, reproduced from
Paper II, lists the stellar parameters
of our sample of M71 stars.

\subsection{Equivalent Widths}

The search for absorption features present in our HIRES data and the
measurement of their equivalent width (\ew) was done automatically with
a FORTRAN code, EWDET, developed for this project. Details of this code
and its features are described in Paper II.
The line list identified and measured by EWDET is then correlated 
to the list of suitable unblended lines with atomic parameters 
to specifically identify the different atomic lines. 
The list of unblended atomic lines was created by inspection of the spectra
of M71 stars, as well as the online Solar spectrum taken with the FTS
at the National Solar Observatory of
Wallace, Hinkle \& Livingston (1998) and
the set of Solar line identifications of Moore, Minnaert \& Houtgast (1966).

In Paper II, we derived the $\lambda D-$\ew\ relation of the Fe I 
lines of ``the weak line set'' (Fe I lines within two sigma levels of the 
$\lambda D-$\ew\ fit, \ew\ $<$ 60 m\AA, and errors less than a third of the 
\ew).
We used these $\lambda D-$\ew\ relations to determine ``the good line set''
(lines with errors less than a third of the \ew\ and with \ew\
computed from the derived $\lambda D-$\ew\ relations).
The \ew\ of the lines presented in this paper are also determined using the 
fit to the $\lambda D-$\ew\ relation of the Fe I lines of ``the good line set'',
except for the C I, O I and Ca I lines, and for the elements that 
show hyperfine 
structure splitting (Sc II, V I, Mn I, Co I, Cu I, and Ba II). 
The equivalent widths of the C I and O I lines were measured 
by hand, since thermal motions become important at the low atomic weights
of these elements and 
the $\lambda D-$\ew\ relations derived for Fe I lines may no longer be valid. 
For Ca I lines and the lines of elements that show hyperfine structure splitting, 
we used the equivalent widths measured automatically by EWDET,
but did not force them to fit the Fe I $\lambda D-$\ew\ relationship due to
the probable different broadening mechanisms. Many of the Ca I lines 
were strong 
enough to be on the damping part of the curve of growth.
The \ew\ used in the abundance analysis are listed in Table~\ref{tab2} 
(available electronically), which also includes the \ew\ for the Fe I and Fe II 
lines used in Paper II.

\subsection{Transition Probabilities}

Transition probabilities for the lines used in this analysis 
were obtained from the NIST Atomic Spectra Database (NIST Standard 
Reference Database \#78, see \citet{wei69,mar88,fuh88,wei96}) when possible. 
Nearly 80\% of the lines selected as suitable from the 
HIRES spectra have transition probabilities from the NIST database.
For the remaining lines the $gf$ values come from the inverted solar
analysis of \citet{the89,the90}, corrected by the factors 
listed in Table~\ref{tab3} which are needed to place both sets of 
transition probabilities onto the same scale.
The correction factor was computed as the mean difference 
in log($gf$) between the NIST and solar values for the lines in common,
which number is given in column 2 of Table~\ref{tab3}. 
Elements not listed in Table~\ref{tab3}
have transition probabilities from the NIST database for
all their lines utilized  here.

Six elements show hyperfine structure splitting (Sc II, V I, Mn I, Co I, Cu I, 
and Ba II). The corresponding hyperfine structure constants were taken 
from \citet{pro00}.

\subsection{Damping Constants}

Most of the Na I and Ca I lines are strong enough for damping effects 
to be important.
For Na I the interaction constants, $C_{6}$, of the van de Waals broadening were
taken from the solar analysis of \citet{bau98}. \citet{smi81} studied 
collisional broadening of 17 Ca I lines. Comparing their experimental results
and the predicted values of $C_{6}$ 
obtained using the Uns\"{o}ld approximation, 
we found that the experimental $C_{6}$ are about 10 times larger than the
 Uns\"{o}ld $C_{6}$. Thus for the Ca I we used the experimental $C_{6}$
from \citet{smi81} when available, otherwise we use 10 times the Uns\"{o}ld 
approximation. The empirical values of $C_{6}$ for Al I and Mg I from 
\citet{bau96} and \citet{zha98}, respectively, are also used. 
We used 4 times times the Uns\"{o}ld approximation for those Al I lines without
empirical damping constants.
For the lines of all other ions we set $C_{6}$ to be
twice the Uns\"{o}ld approximation as was done
in Paper II for the Fe I lines following \citet{hol91}.

\subsection{Solar Abundances}

We need to establish the solar abundances corresponding to our adopted set of
$gf$ values and damping constants.
Solar abundance ratios were computed using our compilation of atomic parameters, 
the Kurucz model atmosphere for the Sun \citep{kur93b} and 
the list of equivalent widths from \citet{moo66}.
The results are listed in Table~\ref{tab4}. 
The O abundance from the permitted lines are corrected  by a factor of 
0.35 dex (see below). 
There is a general agreement with the meteoric 
solar abundance ratios from \citet{and89}. 
The difference between our solar abundances and the meteoric solar abundances 
from \citet{and89} is listed in column 5 of Table~\ref{tab4}.
This difference is within the standard deviation of our own 
measurements, with the exception of [Ca/Fe], which is our most deviant 
abundance ratio. The difference we found is almost the same as the correction
factor applied to the solar $gf$ values, listed in Table~\ref{tab3}.
We use these solar abundance ratios, derived from our choice of
atomic line parameters, to compute the abundance 
ratios for our sample of M71 stars.

\subsection{Non-LTE effects \label{section_nonlte}}

The non-LTE treatment of the oxygen permitted lines is discussed in  
\S\ref{abun}. The K I resonance lines are strongly affected by non-LTE effects
in the Sun \citep{del75}. \citet{tak01} carried out statistical equilibrium 
calculations for the K I line 7699 \AA, 
the only line used in our analysis, for metal poor stars.
We applied a non-LTE correction to our results of K abundance  
following \citet{tak01}. The smallest correction applied was  $-$0.12 dex 
for the cool giant stars, and the largest correction was $-$0.7 dex for the
HB stars. Without these corrections, a very strong dependence of
K abundance on \teff\ (equivalent to luminosity) was seen.

The aluminium lines are also affected by non-LTE in metal poor stars
\citep{bau97}. Unfortunately, the statistical equilibrium calculations of
\citet{bau97} included only dwarf stars. Their non-LTE corrections increase
with decreasing metallicity, and are larger for the 3961 \AA \ line than for
the 6697 \AA\ doublet used in our analysis. Al and all other elements were
treated assuming LTE.

\section{ABUNDANCE ANALYSIS \label{abun}}

Given the stellar parameters from Paper I, we determined the abundances
using the equivalent widths obtained as described above.
The abundance analysis is done using a current version of the LTE 
spectral synthesis program MOOG \citep{sne73}. 
We employ the grid of stellar atmospheres from \citet{kur93b} to compute 
the abundances of C, O, Na, Mg, Al, Si, K, Ca, Sc, Ti, V, Cr, Mn, Co, Ni, 
Cu, Zn, Y, Zr, Ba, La, and Eu using the four stellar atmosphere 
models with the closest \teff\ and \grav\ to each star's parameters.
The abundances were interpolated using results from the closest stellar model 
atmospheres to the appropriate \teff\ and \grav\ for each star. 

We also determine the abundance of Li using a synthesis of the spectra in 
the area of the doublet at 6707\AA, which, since it is not well resolved,
is considered as only one line in Table~\ref{tab5a}.
The Li abundance is given as log~$\epsilon$(Li)=log($N_{Li}/N_{H}$)+12.0,
where $N$ is number of atoms.
In this case, we used the stellar model 
atmosphere from \citet{kur93b} with the closest \teff\ and \grav\ to each 
star's parameters. 

The abundance ratios, with the exception of [C/Fe], [O/Fe], [Si/Fe] and [Zn/Fe] 
are computed using the iron abundance from Fe I lines of Paper II
as slightly updated in Table~\ref{tab1}, 
and our solar abundance ratios from Table~\ref{tab4}. 
Given their high excitation potentials, the abundance ratios for the
C I, Si I, and Zn I lines were computed using the [Fe/H] from Fe II lines.
In the \teff\ range of our sample of stars, most of the iron is in the 
form of Fe II and most of the oxygen is in the form of O I, so both 
species behave similarly for small changes in the atmospheric parameters.
For this reason, we computed the abundance ratio of O using the 
Fe II lines as well.
The computed abundance ratios are listed in Tables ~\ref{tab5a} - ~\ref{tab5e}. 

\begin{figure}
\epsscale{0.7}
\plotone{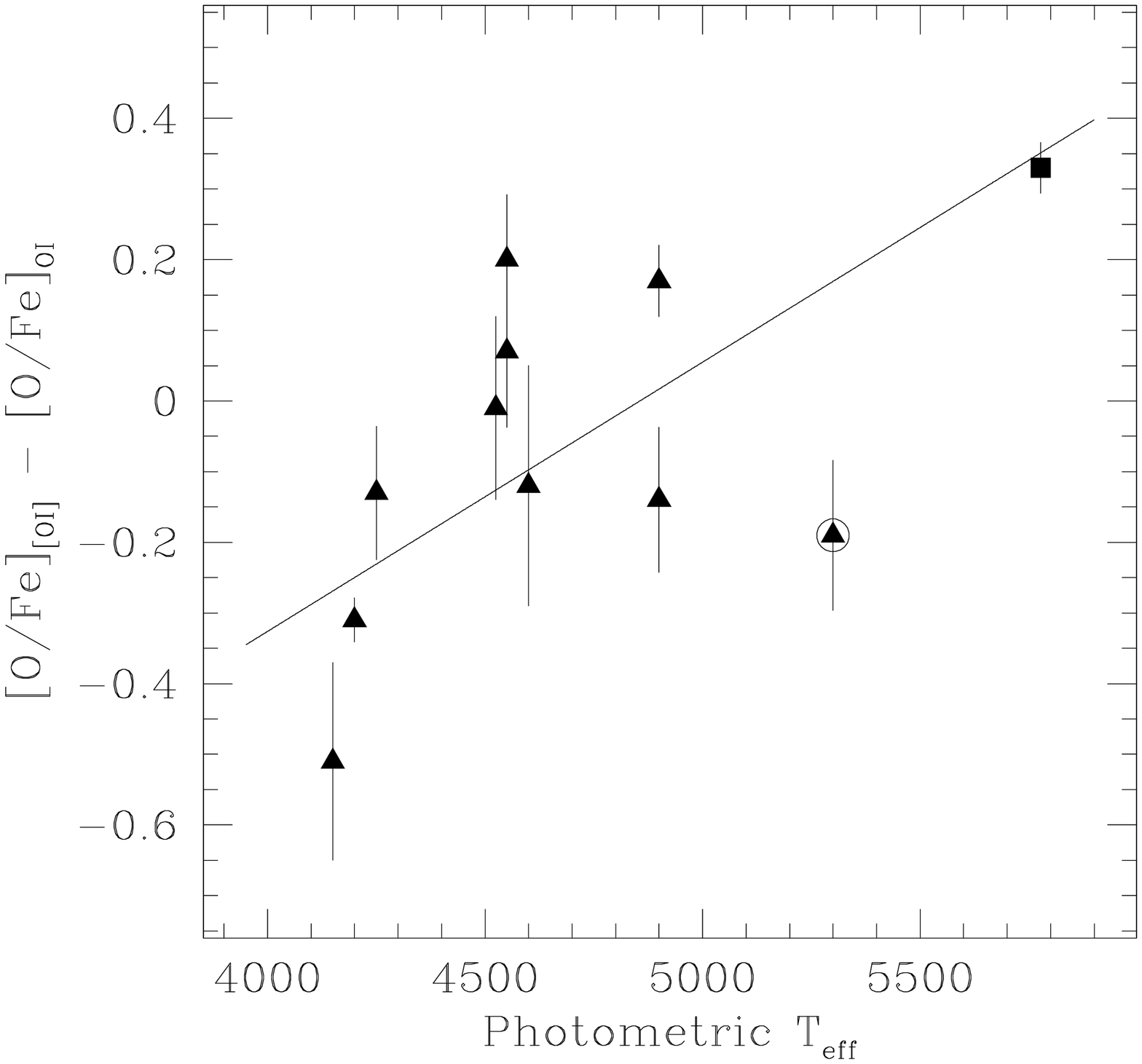}
\figcaption[ox_teff.ps]{The difference between the oxygen abundance ratio from the forbidden
and permitted lines is shown as a function of \teff. 
The solid line is a linear fit weighted by the errors.
The O abundances subsequently deduced from the permitted lines are 
corrected  by 
the linear fit shown here.
The RHB stars are marked with  open circles, while the position of the
Sun is indicated by a filled square.
\label{ox_teff}}
\end{figure}

There are 11 stars with equivalent widths measured for both the two forbidden and 
the permitted O lines. The difference of the oxygen abundance ratio from
forbidden and permitted lines for those 11 stars and the Sun 
is plotted against \teff\ in Figure~\ref{ox_teff}. 
A clear trend with \teff\ is observed, which may come from the different 
excitation potential of the forbidden and the permitted lines or from 
non-LTE effects on the permitted lines. 
We tried applying the non-LTE corrections suggested by \citet{gra99} and
by \citet{tak02} to
the permitted lines, but the observed \teff\ trend becomes even steeper. 
\citet{mel01} and \citet{lam02}, among others, discuss the validity of 
the different oxygen abundance indicators. 
Since the forbidden lines are usually considered to give more 
reliable abundances \citep[but see ][]{isr01,all01}, we corrected the 
abundance ratio from the permitted lines by the amount given by the least 
squares fit shown in Figure~\ref{ox_teff}. 
The final [O/Fe] listed in Table~\ref{tab5a} is the average for each star
of the results from the forbidden and the corrected permitted lines.
Note that a correction of 0.35 dex, which corresponds to the 
correction at
the temperature of the Sun, was applied to the abundances deduced 
from the permitted lines of O I in the Sun to compute its 
[O/Fe] in the same way as for our M71 sample of stars.

\begin{figure}
\epsscale{0.7}
\plotone{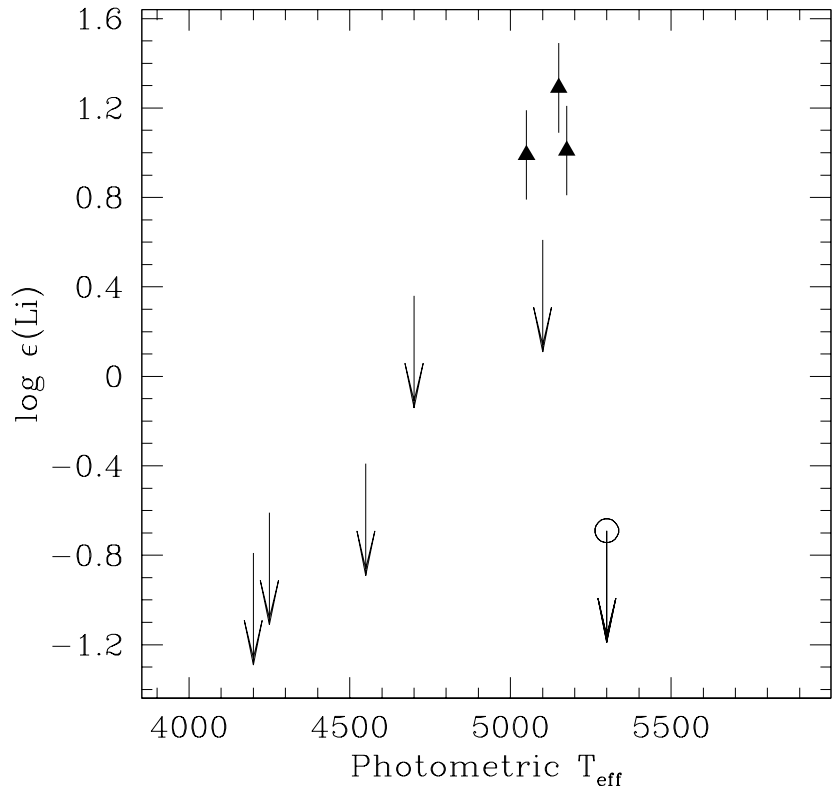}
\figcaption[li1.ps]{log $\epsilon$(Li) against \teff.
The RHB stars are marked by open circles.
Arrows represent upper limits for the strength of the Li I line.
\label{li1}}
\end{figure}

\begin{figure}
\epsscale{0.7}
\plotone{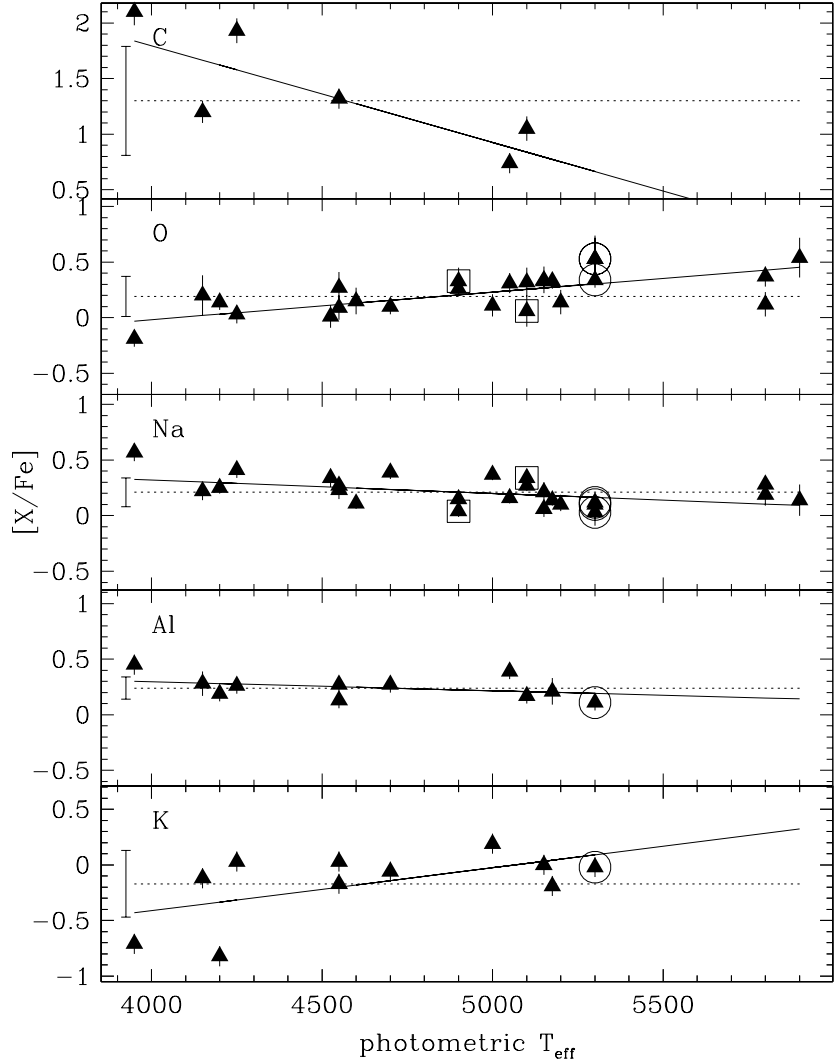}
\figcaption[light.ps]{Abundance ratios of C, O, Na, Al, and K with
respect to Fe against \teff.
The solid line is a linear fit weighted by the errors.
The dashed line indicates the mean abundance ratio with its respective error plotted
as an error bar at 3925 K.
The RHB stars are marked by open circles.
A non-LTE correction has been applied to the O I permitted lines and the K I
line.
The C abundances determined from C I lines 
in the cooler M71 stars are believed
to be spurious due to contamination by lines from the red system of CN 
(see text).
Stars 1--60 and 2--160, part of whose spectra are shown in Figure~\ref{na_o_spec}, 
are marked with open squares in the [O/Fe] and [Na/Fe] panels.
\label{light}}
\end{figure}

\begin{figure}
\epsscale{0.7}
\plotone{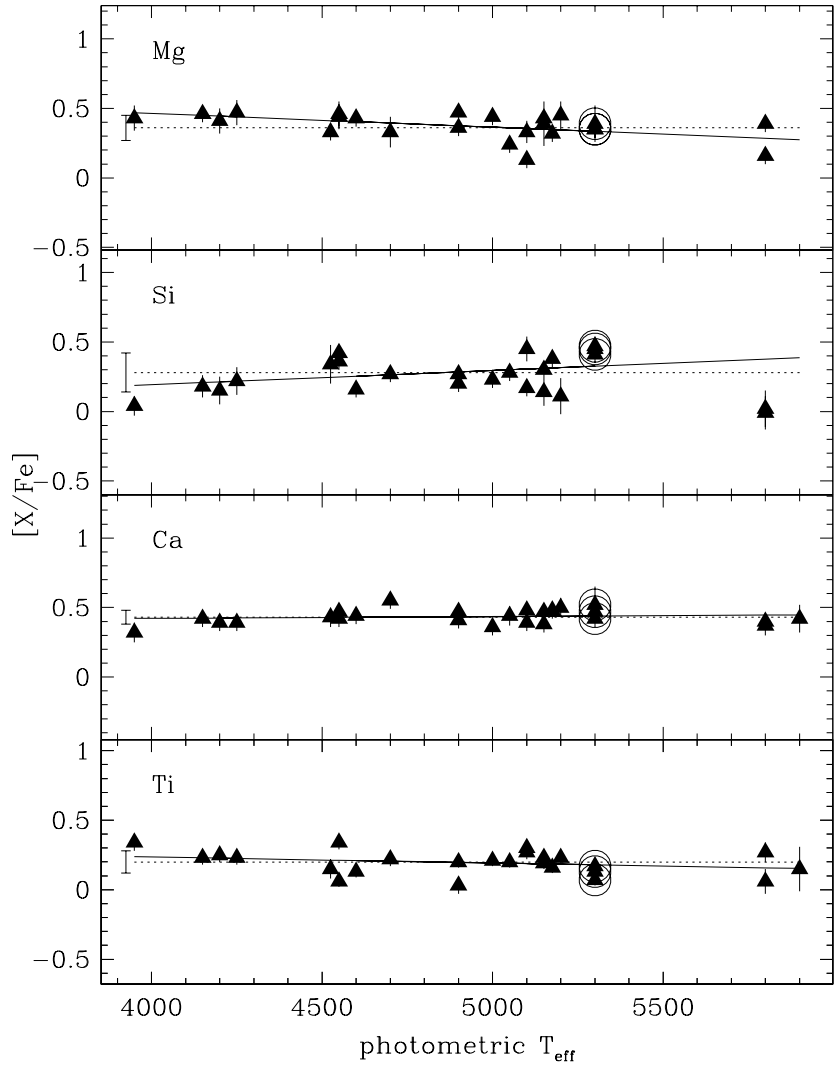}
\figcaption[alpha.ps]{Abundance ratios of the $\alpha-$elements 
Mg, Si, Ca and Ti
with respect to Fe against \teff.
The symbols are the same as in Figure~\ref{light}.  
\label{alpha}}
\end{figure}

\begin{figure}
\epsscale{0.7}
\plotone{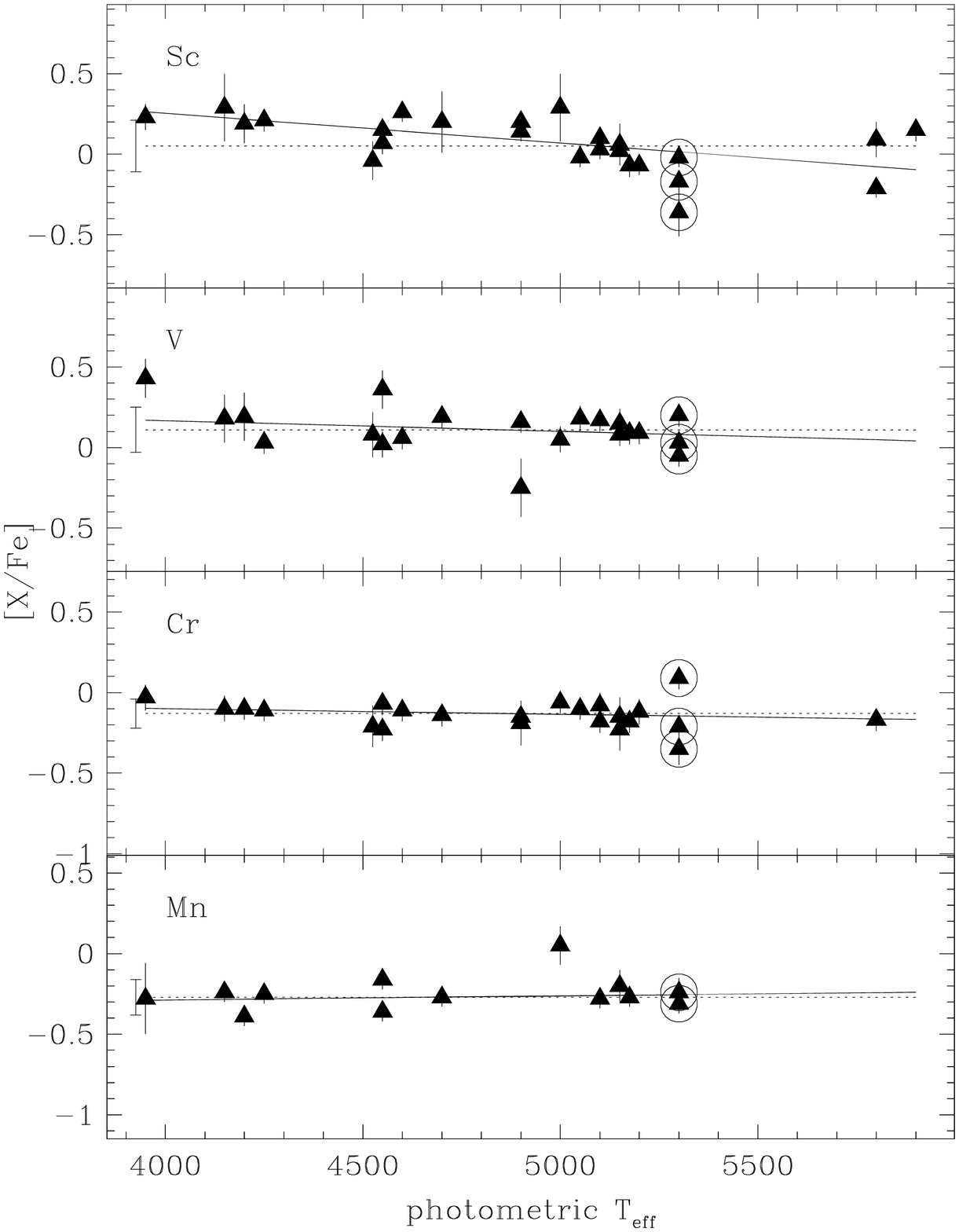}
\figcaption[ironp1.ps]{Abundance ratios of the iron peak elements 
Sc, V, Cr, and Mn
with respect to Fe against \teff.
The symbols are the same as in Figure~\ref{light}.
\label{ironp1}}
\end{figure}

\begin{figure}
\epsscale{0.7}
\plotone{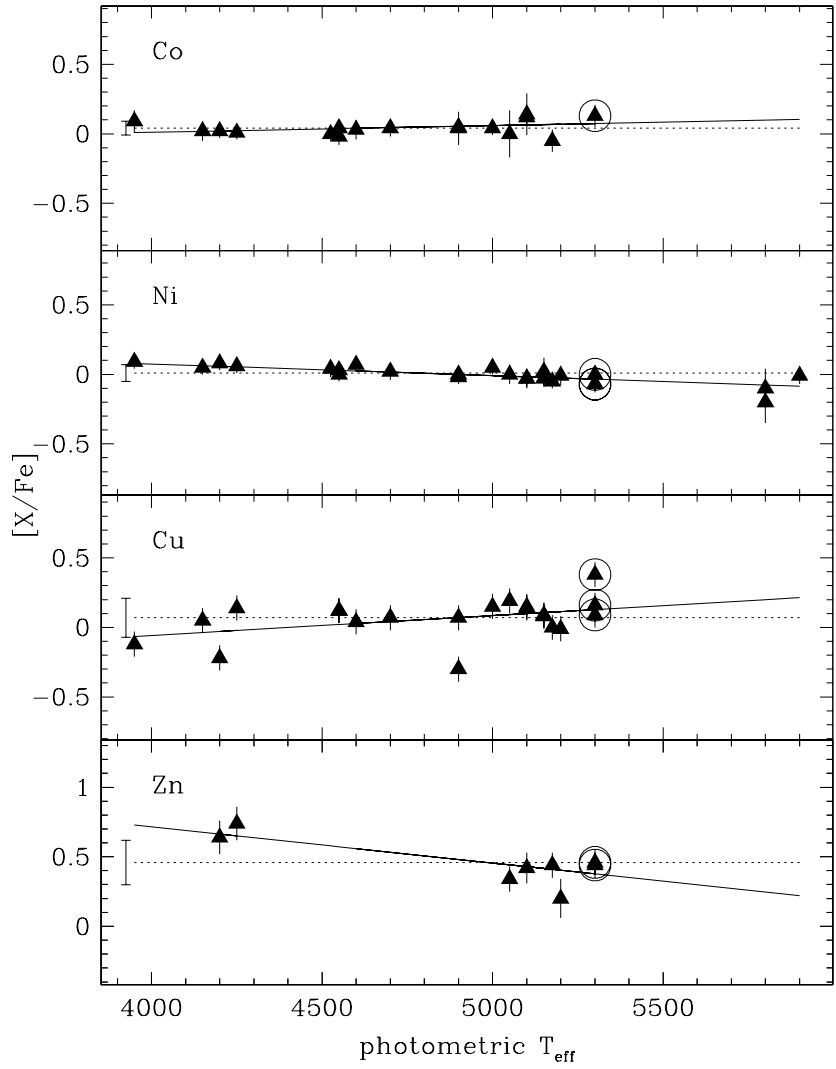}
\figcaption[ironp2.ps]{Abundance ratios of the iron peak elements 
Co, Ni, Cu, and Zn
with respect to Fe against \teff.
The symbols are the same as in Figure~\ref{light}.
\label{ironp2}}
\end{figure}

\begin{figure}
\epsscale{0.7}
\plotone{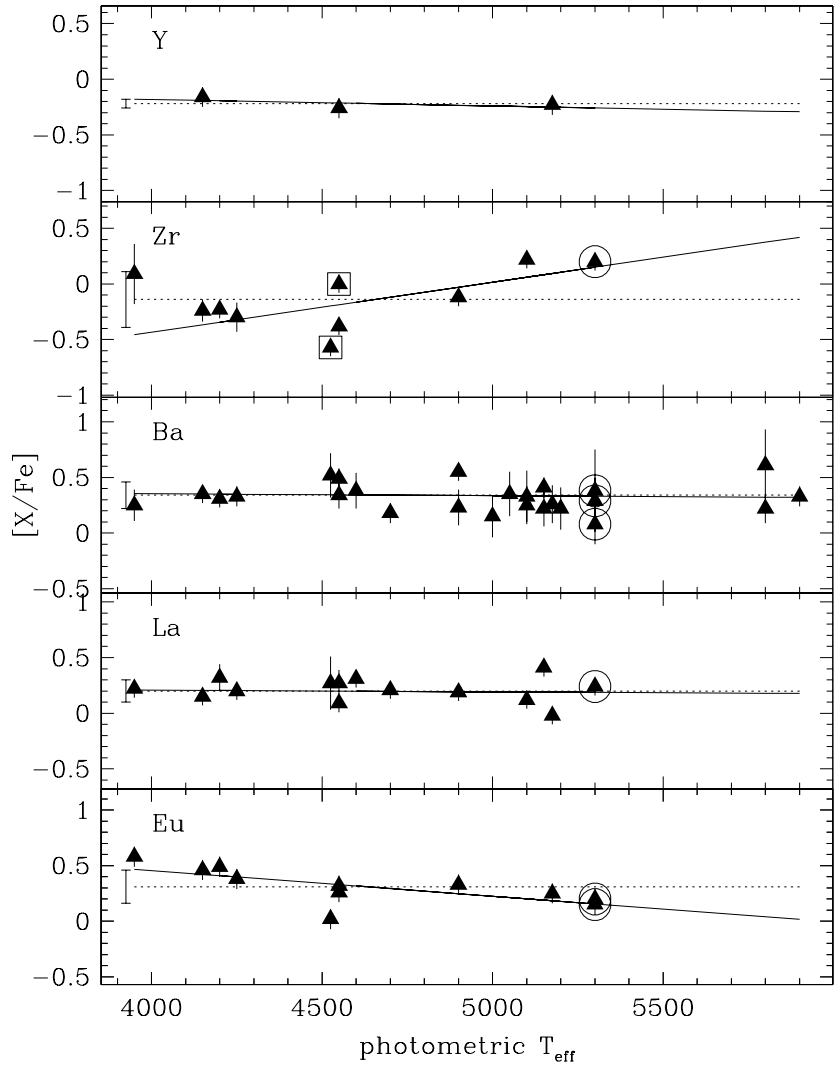}
\figcaption[neutron.ps]{Abundance ratios of the neutron capture elements 
Y, Zr, Ba, La, and Eu
with respect to Fe against \teff.
The symbols are the same as in Figure~\ref{light}.
Stars 1--56 and 1--81, part of whose spectra are shown in Figure~\ref{zr_spec}, 
are marked with open squares in the [Zr/Fe] panel.
\label{neutron}}
\end{figure}

The abundance ratios (absolute abundance for Li) for each star in our M71 sample
are plotted against the photometric \teff\ in 
Figures~\ref{li1} to ~\ref{neutron}. 
The error bars shown in Figures~\ref{li1} to~\ref{neutron} correspond to the 
standard deviation of 
results of different atomic lines divided by the square root 
of the number of lines used for each star.
The solid line, shown in Figures~\ref{light} to~\ref{neutron}, is a linear 
fit weighted 
by the errors of the respective abundance ratio versus \teff.
The dashed line shown in these figures indicates the mean 
abundance ratio and its respective error plotted as an error bar at 3925 K.
The error in the mean abundance ratio corresponds to the standard deviation 
within our sample of stars divided by the square root of the number of 
stars for which an abundance was derived for that ion.

We estimate the sensitivity of the abundances with respect to small changes 
in the equivalent widths (synthesis for Li) and the stellar parameters 
in four cases 4000/1.0/1.4, 4250/1.0/1.4, 5000/2.5/1.0 and 
5500/4.0/0.6, where the three numbers correspond to \teff/\grav/\mtv.
The case 4000/1.0/1.4 has been computed only for elements with high excitation 
lines, which are more sensitive at lower temperatures.
The stellar parameters of these cases span the relevant range of 
atmospheric parameters for our M71 sample.
We estimated the error in the \ew\ to be 10\% for all the lines.
The error of the synthesis of the Li doublet is estimated to be $\pm$0.1 dex.
The results are listed in Table~\ref{tab6}, where the range adopted for each 
parameter is representative of its uncertainty.

Because of the high excitation of the C I lines studied here,
this is the ion included in our analysis
whose derived abundance is most sensitive to \teff. 
[Ca/H] also has a sensitive dependence on \teff\ and on \mtv, 
because the Ca I lines are all rather strong and have large 
damping constants.

The mean abundance ratios and their errors are listed in Table~\ref{tab7}. 
The statistical error, $\sigma_{obs}$, corresponds to the standard deviation of
sample of stars divided by the square root of the number of stars and
it is a measure of the scatter of the abundance ratio in the sample of M71 stars.
In order to quantify the abundance ratio variations within our sample of 
M71 stars we have to compare the measure of the scatter with the predicted
error from the stellar parameters and the measurement of the \ew\ (or synthesis
for Li).
We estimated the predicted error, $\sigma_{pred}$, using the following equation:
\begin{eqnarray*}
\sigma^{2}_{pred}([X/Fe]) = 
 & \Delta (X:W_{\lambda})^{2}/N_{lines}(X)+\Delta (Fe:W_{\lambda})^{2}/N_{lines}(Fe) + \\
 & [\Delta (X:T_{eff}) - \Delta (Fe:T_{eff})]^{2} + \\
 & [\Delta (X:{\rm log}(g)) - \Delta (Fe:{\rm log}(g))]^{2} + \\
 & [\Delta (X:\xi)  - \Delta (Fe:\xi)]^{2} +  \\
 & [\Delta (X:{\rm [Fe/H]}) - \Delta (Fe:{\rm [Fe/H]})]^{2} 
\end{eqnarray*}
where $\Delta(X:$\ew), $\Delta(X:$\teff), $\Delta(X:$\grav), $\Delta(X:$\mtv), 
and $\Delta(X:$\fe) are
listed in columns 2, 3, 4 ,5 and 6 of Table~\ref{tab6}. 
$N_{lines}$ is the number of lines used
to compute the abundances, $X$ denotes the element under consideration,
and $Fe$ denotes either Fe I or Fe II, whichever was used to 
compute the abundance ratio. 
Our $\sigma_{pred}$ ignores covariance among the error terms, which
is discussed in detail by \citet{joh02}.  She shows that these
additional terms are fairly small, and will be even smaller in our case, 
as we have determined \grav\ using isochrones rather than through
ionization equilibria (see Paper I).
The general small trends seen in Figures~\ref{li1} to \ref{neutron} 
of [X/Fe] slightly increasing toward cooler \teff\
may result from ignoring the covariance terms (see Johnson 2002).

The predicted errors for each ion are listed in column 4 of Table~\ref{tab7}.
The maximum abundance trend over the 
relevant \teff\ 
range for each element, $\Delta_{max}$, is also listed
in column 6 of Table~\ref{tab7}.  This parameter, which is not
sensitive to star-to-star scatter abundance variations for stars
at a given evolutionary state,
is the slope of the linear fit of the abundance
ratio vs. \teff\ times the range in \teff; its error is the error in the
slope time the range in \teff\ covered by the sample of stars
in which the ion of interest was observed.  The values of $\Delta_{max}$
for essentially all elements observed are gratifyingly small, 
providing evidence to support many of the assumptions made in the course
of this analysis, such as that of non-LTE.

\begin{figure}
\epsscale{0.7}
\plotone{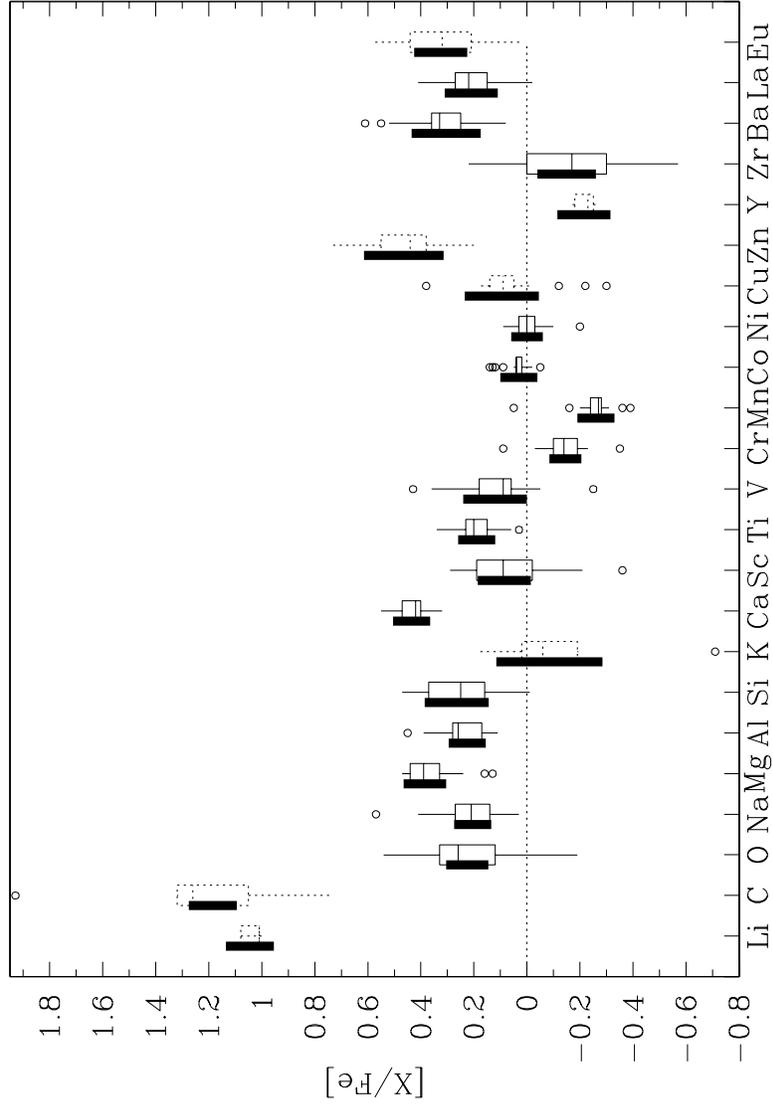}
\figcaption[summ_fig2.ps]
{Summary of abundance ratios. Each abundance ratio is plotted with a box, 
which central horizontal line is the median abundance ratio, the bottom 
and the top shows its inter--quartile range, the vertical lines coming 
out of the box mark the position of the adjacent points of the sample, 
and the outliers are plotted as open circles. Boxes constructed
with dashed lines denote elements where only one line per star was
observed. The thick line on the left 
side of the box is the predicted error (expected for the 
inter--quartile range) which included the dependence on the stellar
parameters and the equivalent width determination.
\label{summ_fig2}}
\end{figure}

A summary of the abundance ratios for our M71 sample
is shown in Figure~\ref{summ_fig2}. 
The results for each element are depicted as
a box whose central horizontal line is the median 
abundance ratio for all the M71 stars included, while the bottom and
the top shows its inter--quartile range, the vertical lines 
coming out of the box mark the
position of the adjacent points of the sample, 
and the outliers are plotted as open circles. 
The  boxes drawn with dotted lines correspond to 
elements with abundances computed from only one line in each star and 
hence are more uncertain.
The thick line on the left side of the box is the predicted 
1$\sigma$ rms error
scaled to correspond to the $\pm$25\% inter--quartile range. 
As seen in Figure~\ref{summ_fig2}, the elements where 
we expect to see star-to-star variations in our M71 sample
are O, Na, Zr, and the special cases of Li and C,
each to be discussed in detail later. 

\section{DISCUSSION}

\subsection{Fe-peak elements}

The abundance ratios of [Sc/Fe], [V/Fe], [Cr/Fe], [Mn/Fe], [Co/Fe], and [Ni/Fe] 
follow the behavior of iron as expected, showing no significant trend with \teff, 
and less scatter around the mean than the predicted error. 
The mean abundance ratios of Sc ($<$[Sc/Fe]$>$=+0.05$\pm$0.16), V 
($<$[V/Fe]$>$=+0.11$\pm$0.14), and Ni ($<$[Ni/Fe]$>$=+0.01$\pm$0.06) are 
consistent with the earlier results of \citet{sne94}, who analyzed high resolution 
spectra of ten giant stars in M71, obtaining $<$[Sc/Fe]$>$=+0.10$\pm$0.03, 
$<$[V/Fe]$>$=+0.19$\pm$0.04, and $<$[Ni/Fe]$>$=+0.07$\pm$0.04. 
Our abundance ratios of the iron peak elements are also consistent with
the results of \citet{lee87}.

The Zn I line at 6362.3 \AA\ is definitely present in the best of the        
spectra of the M71 giants, but                                                
is somewhat blended with the much stronger Fe I line at 6362.9.               
In addition,                                                                  
the continuum there is depressed due to a broad auto-ionization Ca I feature. 
The rather high abundance of Zn                                               
we deduce must thus be regarded as quite uncertain                            
until a full spectral synthesis                                               
of this region becomes available.                                             
                                                                              
\subsection{Neutron capture elements}

We have detected lines of the neutron capture elements Y, Zr, Ba, La and Eu. 
\citet{cam82} and \citet{kap89} analyzed the solar system 
meteoritic abundances of neutron capture elements to yield accurate 
breakdowns into $r$- and $s$-process parts for each isotope, 
which have been summed into fractions for each element by \citet{bur00}.  
At [Fe/H] $<-$2.0, as reviewed by \citet{sne01},
Zr, Ba, and La are neutron capture element synthesized through $s-$process
reactions that occur mainly in low mass asymptotic giant branch (AGB) stars,
while Eu is exclusively an $r-$process element.

\begin{figure}
\epsscale{0.7}
\plotone{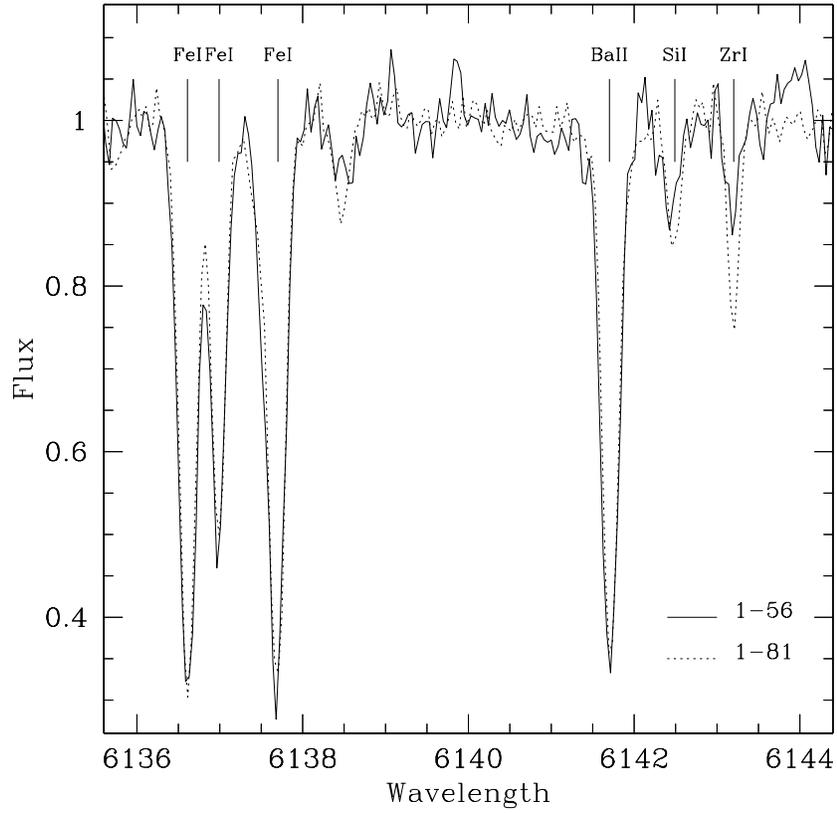}
\figcaption[zr1_spec.ps]{Comparison of the strength of the
strongest Zr I line included in our study between two
stars of similar effective temperatures, 1--56 (4525 K, [Zr/Fe]=--0.57)
and  1--81 (4550 K, [Zr/Fe]=0.00). The scatter shown by [Zr/Fe]
might be due to real abundance variations among stars of different \teff.  
\label{zr_spec}}
\end{figure}

The abundance ratios of the neutron capture elements, Y, Zr, Ba, La, and Eu, 
show no significant trend with \teff, and less scatter around 
the mean than the predicted error, except for [Zr/Fe].
In Figure ~\ref{zr_spec}, we show the spectra  for two stars 
of similar \teff\ and different [Zr/Fe], 1--56 (4525 K, [Zr/Fe]=--0.57) and 
1--81 (4550 K, [Zr/Fe]=0.00) in the region of the Zr line at 6143 \AA, 
which is the strongest Zr I line included in our study.
In the spectral range illustrated, there are also two Fe I lines, one Ba II
line and one Si I line, whose strengths are similar in both stars. 
It is possible but not certain that the difference in strength of 
the Zr line is due to star-to-star abundance variations.
\citet{lee87} analyzed four Zr lines in five bright giant M71 stars
to obtain [Zr/Fe] $\sim 0.0$ dex.

The abundances of Ba, La, and Eu are overabundant relative to Fe, as 
is seen in other clusters (see below). 
The mean [Ba/Eu] ratio of +0.03 is consistent with values 
observed in halo stars of similar [Fe/H] \citep{bur00,gra94}. 

\subsection{$\alpha-$elements}

We find that the $\alpha-$elements Mg, Ca, Si and Ti are overabundant relative 
to Fe. 
Our mean $<$[Ti/Fe]$>$=+0.20$\pm$0.08 and $<$[Si/Fe]$>$=+0.28$\pm$0.14 are 
similar to the results of \citet{sne94} for ten M71 giant stars 
($<$[Ti/Fe]$>$=+0.48$\pm$0.04, $<$[Si/Fe]$>$=+0.31$\pm$0.04), and
also similar to the abundance ratios provided by \citet{lee87}.
Our $<$[Ca/Fe]$>$=+0.43$\pm$0.05 is higher than the value of \citet{sne94} 
($<$[Ca/Fe]$>$=+0.13$\pm$0.03), but similar to the abundance ratio 
of +0.58 found by \citet{lee87}.
The $\alpha-$element abundance ratios show no significant trend with \teff, 
and low scatter around the mean.

[Mg/Fe] is know to vary among bright giant stars in some metal poor globular 
clusters.
In NGC 6752 \citep{gra01}, M13 \citep{kra93,she96} and M15 \citep{sne97}, 
[Mg/Fe] shows a star-to-star range in abundance of about 1.5, 1.2, 
and 1.0 dex respectively.
Our comparison between the observed scatter and the predicted error
of [Mg/Fe] given in Table~\ref{tab7} indicates no sign of 
star-to-star variation of magnesium in M71.

\subsection{Sodium and Oxygen \label{section_o}}

The oxygen abundance ratios in our sample of stars in M71 behave differently 
than the abundance ratios of all other elements included in this paper. 
The scatter in [O/Fe] versus \teff\ shown in Figure~\ref{light} 
strongly suggests that O shows star-to-star variations within the 
M71 sample.
Furthermore, the observed scatter for [O/Fe] given in
Table~\ref{tab7} is larger than the
respective predicted error which include the effects of uncertainties in the
determination of the the stellar parameters and
in the equivalent width measurements.
To a lesser extent, the Na abundances behave similarly, as shown
in Figure~\ref{light}, and the observed scatter for Na is slightly
larger than the value $\sigma_{pred}$ given in Table~\ref{tab7}.

\begin{figure}
\epsscale{0.7}
\plotone{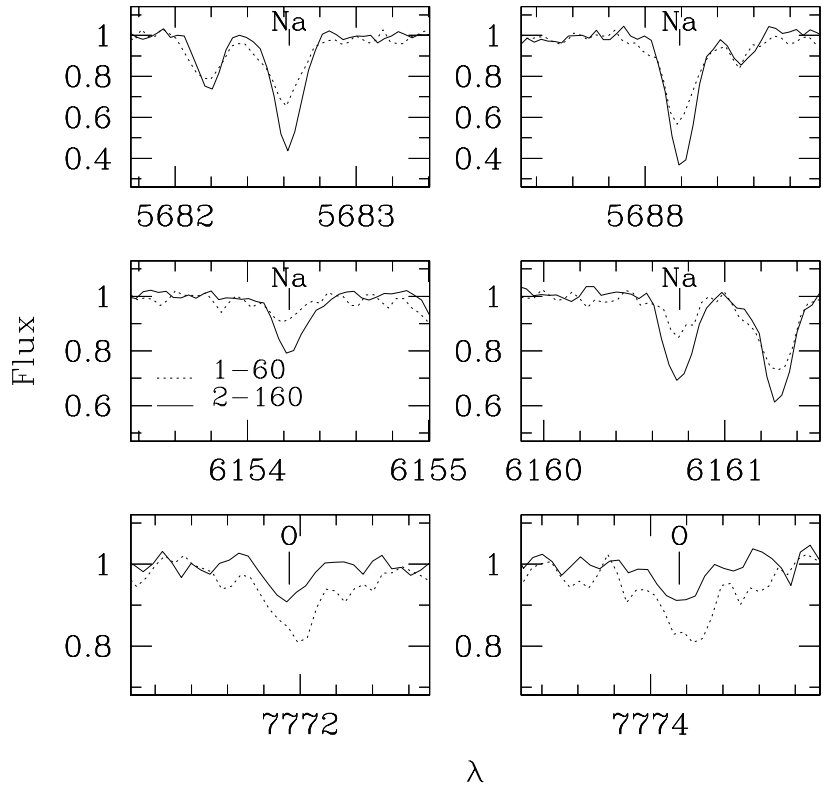}
\figcaption[na_o_spec.ps]{Comparison of the strength of four 
Na I and two O I lines between two 
stars of similar effective temperatures, 1--60 (4900 K, [Na/Fe]=+0.04, [O/Fe]=+0.33) 
and  2--160 (5100 K, [Na/Fe]=+0.34, [O/Fe]=+0.06). The scatter shown by [Na/Fe] and
[O/Fe] is due to real abundance variations among stars of different \teff.  
\label{na_o_spec}}
\end{figure}

In Figure~\ref{na_o_spec} we compare the strength of the Na I and 
O I lines between two stars of similar stellar parameters.
The star with a high [O/Fe] in our sample (1--60) has a low [Na/Fe] and 
the star with a low [O/Fe] (2--160) has an intermediate [Na/Fe].
These two stars are marked with open squares in Figure~\ref{light}.
Note that both of these stars are red giants fainter than the HB,
with $M_V = +1.4$ for the fainter, Na-richer star.
This figure demonstrates 
that the higher scatter seen in [Na/Fe] and [O/Fe] is due to star-to-star 
abundance variations and that in the case of this specific pair of
stars, Na and O are anti-correlated, as was first observed by
\citet{pet80} in M13.

The non-LTE corrections for the infrared OI triplet are not accurately
know (see Sec.~\ref{abun}). However, assuming that they are 
monotonically dependent on ~\teff, the rms in the non-LTE correction 
cannot introduce a star-to-star scatter in the O abundance nor
the Na-O anticorrelation we observe.

\begin{figure}
\epsscale{0.7}
\plotone{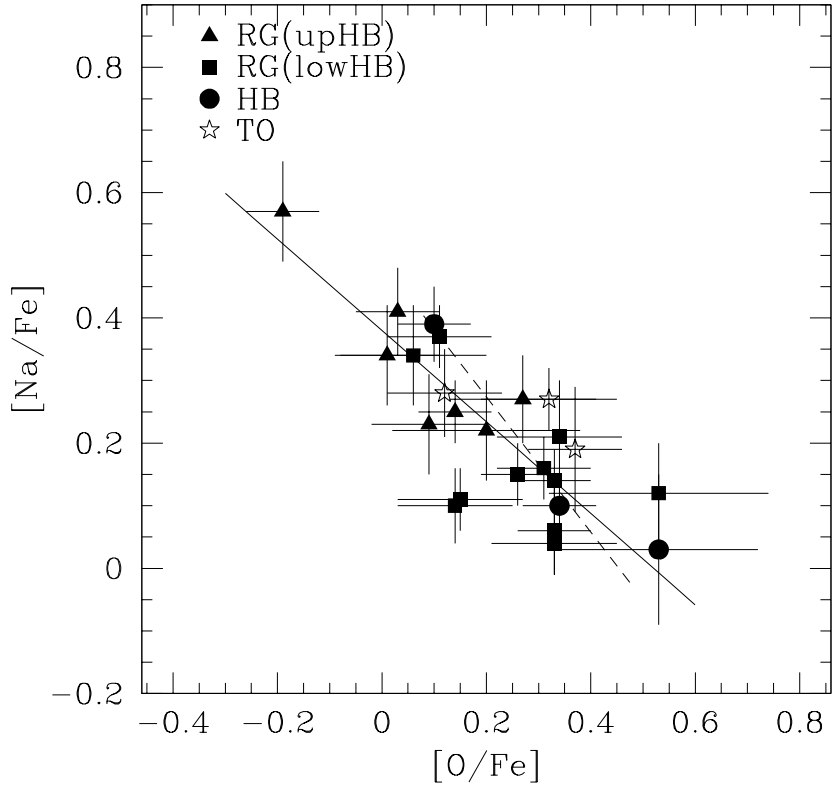}
\figcaption[na_o_m71.ps]{[Na/Fe] against [O/Fe] for M71 stars. 
Filled triangles are 
RG stars brighter than the HB, filled squares are RG stars fainter 
than the HB,  filled circles are HB stars and ``stars'' are stars 
near the main sequence turn off.
The solid line represents the least squares linear fit to our data in M71.
The dashed line corresponds to the Na--O anti-correlation present in 
M4 from the analysis of \citet{iva99}.
\label{na_o_m71}}
\end{figure}

To explore the presence of an anti-correlation between [Na/Fe] and [O/Fe]
within our M71 sample as a whole,
we construct Figure~\ref{na_o_m71},
which presents the Na versus O abundance diagram
for our sample in M71.
Our data are indicated by filled symbols, where triangles are red giants 
brighter than the HB, circles are HB stars, squares are red giants 
fainter than the HB and the stars near the main sequence turnoff are 
denoted by ``stars''. 
We find that Na and O are anti-correlated in our sample of M71 stars.
The fit weighted by the error of Na vs. O, plotted as a solid line
in Figure~\ref{na_o_m71}, is statistically significant
at a 2$\sigma$ confidence level.
Star-to-star Na variations 
and anti-correlation between Na and O abundances extend well beyond
$M_V = +1.4$, and include the small 
sample of stars near the main sequence turnoff, where this anti-correlation
has the same form as among the more luminous stars and
is highly statistically significant at a level exceeding 4$\sigma$.

\begin{figure}
\epsscale{0.7}
\plotone{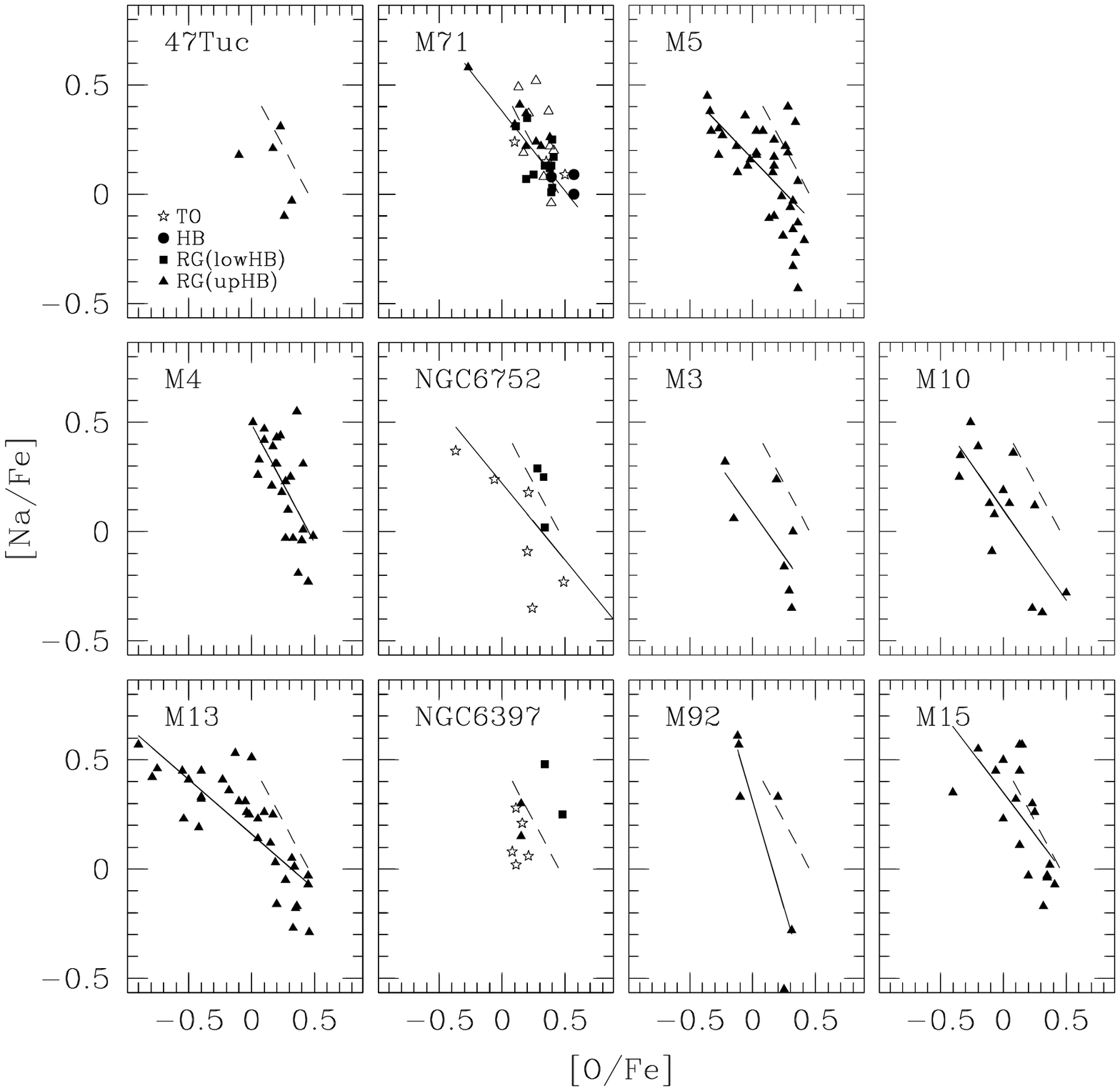}
\figcaption[na_o.ps]{[Na/Fe] against [O/Fe] for
M71 and for other globular clusters from
the literature. Filled symbols are the same as Figure~\ref{na_o_m71}.
The literature determinations are: 47 Tuc \citep{bro90,bro92,nor95}, 
M5 \citep{iva01,she96,sne92}, M4 \citep{iva99}, NGC 6752 \citep{gra01}, 
M3 \citep{kra93}, M10 \citep{kra95}, M13 \citep{kra93,she96}, NGC 6397 
\citep{cas00,gra01}, M92 \citep{sne92}, and M15 \citep{sne97}.
Open triangles are bright red giants in M71 from the abundance 
analysis of \citet{sne94}.
The solid lines represent the least squares linear fits of the
data from the literature for each cluster.
They only shown for those globular clusters where the slope we derive
is significant at the 2$\sigma$ level. The dashed line corresponds
to the anti-correlation observed in M4 from \citet{iva99}, 
shown as a fiducial line in each panel.  \label{na_o}}
\end{figure}

In view of the large sample of bright RGB stars studied in M4 by 
\citet{iva99}, we adopt their results for the observed anti-correlation
between Na and O among red giants in this cluster to provide a 
fiducial line for visual comparisons in the relevant figures. 
The anti-correlation found from their sample is indicated as a dashed line 
in Figure~\ref{na_o_m71}, as well as in the panels of Figure~\ref{na_o},
to be discussed next.
To within the errors, the Na/O anti-correlation we find in M71
agrees with that of M4, within a 2$\sigma$ confidence level.

In Figure~\ref{na_o}, we compare the determinations of Na and O abundance 
ratios that exist in the literature for metal poor globular clusters, 
47 Tuc \citep{bro90,bro92,nor95}, M71 \citep[][, this paper]{sne94},
M5 \citep{iva01,she96,sne92}, M4 \citep{iva99}, NGC 6752 \citep{gra01},
M3 \citep{kra93}, M10 \citep{kra95}, M13 \citep{kra93,she96}, 
NGC 6397 \citep{cas00,gra01}, M92 \citep{sne92}, and M15 \citep{sne97}.
The symbols are the same as in Figure ~\ref{na_o_m71}. 
Also included in this figure are
the earlier results for ten bright giant 
stars in M71 from \citet{sne94}, shown in open triangles.
All their stars are brighter than the HB and  
behave similarly to our red giants brighter than the HB. 
Our observed range in [Na/Fe] is similar to the range observed by 
\citet{sne94}, but our range in [O/Fe] is twice as big. 

For each globular cluster depicted in Figure~\ref{na_o},
the solid line represent the least squares linear fit of the 
data from the literature\footnote{Only for M71 do we use a fit weighted
by the errors of each abundance determination; for all other clusters,
the errors in the abundance for each star are assumed constant.}.
It is only shown for those globular clusters where the slope we derive is
significant at the 2$\sigma$ level. The dashed line corresponds 
to the anti-correlation observed in M4 from \citet{iva99}, shown 
as a fiducial line.

At this confidence level, we find Na-O anti-correlations
in M71, M5, M4, NGC 6752, M3, M10, M13, M92 and M15. 
The steepest slope is that of M92, and the flattest slope is
that of M13. But, within the 2$\sigma$ level, all the slopes
are identical.
%
%
No statistically significant global
anti-correlation is detected in 47 Tuc or NGC 6397.

47 Tuc ([Fe/H]$\sim$--0.8), M4 ([Fe/H]$\sim$--1.2) and
NGC 6397 ([Fe/H]$\sim$--2.0) have a similar [Na/Fe] versus [O/Fe] relationship 
as does M71 does in terms of abundance ratio ranges and scatter.   While
the form of the relationship appears to be more or less universal,
Figure~\ref{na_o} suggests that the amplitude of the Na/O anti-correlation
among RGB stars is smallest for the two most metal rich globular clusters shown,
47 Tuc (where the published dataset is very small) and M71, as well as
for NGC 6397.

\subsection{Aluminium}

The abundance of Al is also known to vary from star-to-star
within a globular cluster.
Because the Al doublet at 6697\AA\ is not included in the spectral 
coverage of the primary set of HIRES spectra (see Paper I), it can
be measured only in a subset of the sample of stars studied here. 
Our comparison between $\sigma_{obs}$ and the predicted error 
of [Al/Fe] given in Table~\ref{tab7} indicates that the
scatter for this element abundance ratio is slightly larger to its respective
predicted error.

As discussed in \S\ref{section_nonlte}, Al suffers from non-LTE effects.
We have not adopted any corrections, nor have we applied any to the
set of data from the literature assembled for Al.  We do, however, use the
6696\AA\ doublet, which is less susceptible to non-LTE effects than
is the 3961\AA\ line.

\begin{figure}
\epsscale{0.7}
\plotone{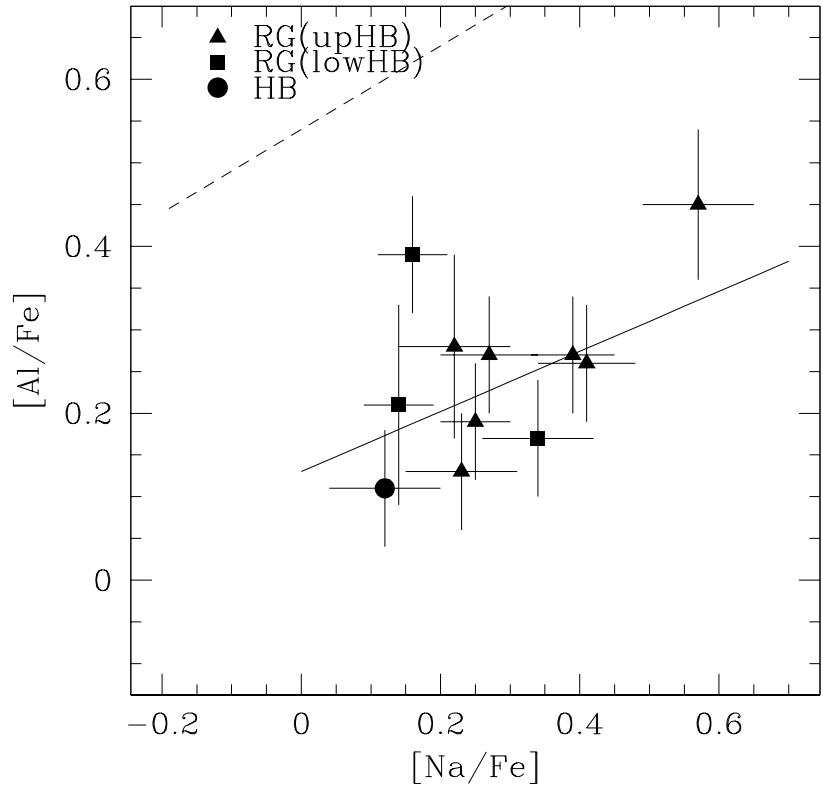}
\figcaption[al_na_m71.ps]{[Na/Fe] against [Al/Fe] for M71 stars. 
Symbols are the same as Figure~\ref{na_o_m71}.
The solid line represents the least squares linear fit to our data in M71.
The dashed line corresponds to the Na--Al correlation present in M4 from the
analysis of \citet{iva99}.  
\label{al_na_m71}}
\end{figure}

We have constructed Figure~\ref{al_na_m71} to explore the presence 
of a correlation between [Na/Fe] and [Al/Fe] in M71, seen in other 
globular clusters. 
The symbols in Figure~\ref{al_na_m71} are the same as in Figure~\ref{na_o_m71}.
We include only our own data in this figure.  A clear Al/Na
correlation is seen which is
statistically significant at a 2$\sigma$ level.  This
correlation extends down to $M_V = +1.8$ mag in M71, where the sample
ends due to the technical issue of the HIRES spectral coverage.
 
\begin{figure}
\epsscale{0.7}
\plotone{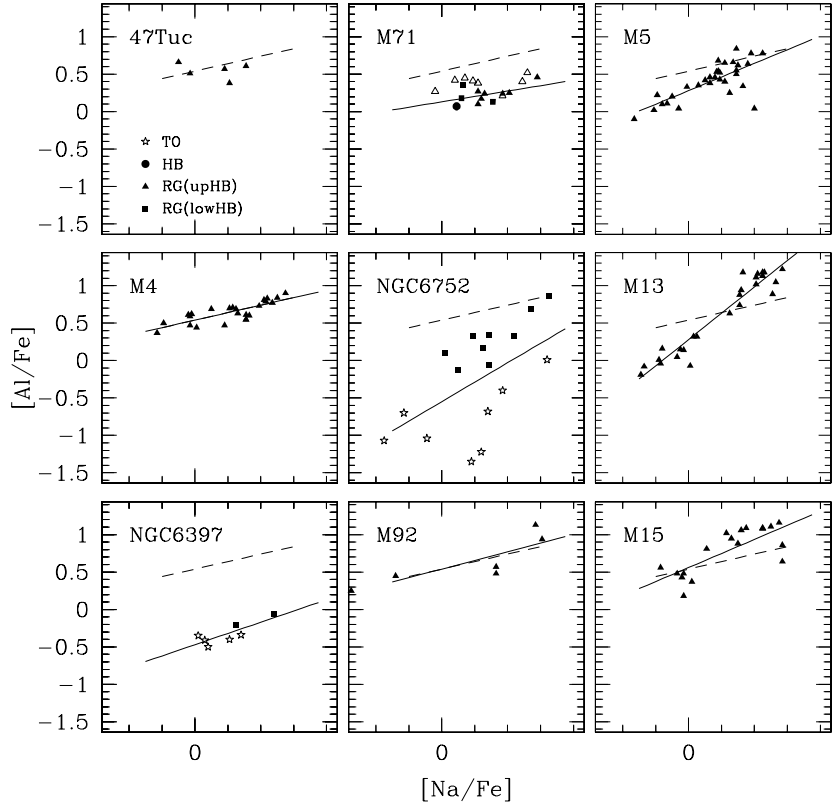}
\figcaption[al_na.ps]{[Na/Fe] against [Al/Fe] for other globular clusters from
the literature. Filled symbols are the same as Figure~\ref{na_o_m71}.
The literature sources used are the same as in Figure~\ref{na_o}.
The solid lines represent the least squares linear fits of the
data from the literature for each cluster.
They only shown for those globular clusters where the slope we derive
is significant at the 2$\sigma$ level. 
The dashed line corresponds to the Na--Al correlation present in M4 from the
analysis of \citet{iva99}, shown as a fiducial line in each panel. 
\label{al_na}}
\end{figure}

In Figure~\ref{al_na}, we compare the determinations of Al/Na abundance 
ratios that exist in the literature for metal poor globular clusters.
We use the same set of references as in \S\ref{section_o}, although
there are fewer stars with measured Al abundances.
The symbols are the same as in Figure~\ref{na_o_m71}.
Again the solid line, representing the least squares linear fit of the
data, is only shown in those globular clusters where the slope we derive is
significant.

At the 2$\sigma$ level, we find Na-Al correlations
in M5, M4, NGC 6752, M13, NGC 6397, M92 and M15, as well as in M71.
At this confidence level, all the slopes,
with the exception of that of M13, are identical.
No anti-correlation is detected in 47 Tuc, where the database is very
sparse.

The differences among the family of linear fits to the Al/Na relationship
for various globular clusters shown in Figure~\ref{al_na} appear
at first sight to be
considerably larger than those shown by the fits to the Na/O anti-correlation
in Figure~\ref{na_o}.  We suggest that these differences 
may not be real, and may arise
from non-LTE effects in Al acting on the different ranges of luminosities
of stars studied in each cluster as well as the particular selection
of Al lines used in each analysis.  The situation in NGC 6752 is particularly
illuminating.  Gratton \etal\ (2001) ascribe the very different 
mean Al abundances deduced for
the subgiants and for the main sequence stars in the cluster
precisely to this issue of ignoring non-LTE in the spectrum of aluminium.

\subsection{Lithium}

Li is a very fragile element and is very easily destroyed in
stars, burning at $T \gtrsim 2.5 \times 10^6$ K.  
\citet{spi82} discovered the presence of Li in
warm halo dwarfs at a constant value (log $\epsilon$(Li) = 2.24)
and suggested that this represents
the primordial Li synthesized in the Big Bang, thus of considerable
importance to cosmology.   Destruction of Li is a measure of
the depth of the surface convection zone, and hence a strong function
of \teff.
\citet{rya01} compile recent
observations for Li in galactic disk and halo stars
and review the Galactic chemical evolution of Li, while
\citet{pin97} reviews the destruction of Li from a theoretical perspective.

We therefore expect Li to be depleted
among the RGB and subgiant stars in globular clusters, although probably
not among the main sequence stars.  
In addition, there is at least one case known
in a globular cluster of the extremely rare class of very
Li-rich stars.  A possible explanation for this star, found 
as a bright RGB star in M3 by 
\citet{kra99}, and similar objects is given by \citet{cha00}.

The Li line is not  included in the spectral coverage
of the primary set of spectra (see Paper I), and hence can
be measured only in a subset of the sample of M71 stars studied here.
We were able to obtain log $\epsilon$(Li) for three giants fainter than the HB,
as well as several upper limits.
The mean log $\epsilon$(Li)  for the detections is 1.10$\pm$0.16
(on the scale H=12.0 dex), 
which is 0.8 dex less than the mean log $\epsilon$(Li) (1.90 $\pm$ 0.42) 
for a sample of 11 halo dwarf stars 
of similar [Fe/H] from the sample of \citet{ful00}, and
is evidence of the strong depletion among the cooler stars
in which Li was detected here. Figure~\ref{li1} illustrates
the pattern of detections and upper limits, which are consistent
with our overall expectations.

\subsection{Carbon}

The analysis of C I lines in cool stars is difficult as the lines are weak, and
their excitation potential is high, $\sim$8.5 eV.  Furthermore, the
C I lines near 7115\AA\ are not included in the spectral coverage
of the primary set of spectra (see Paper I), and hence can
be measured only in a subset of the sample studied here.
We have reliable detections in only 6 stars,
2 lines in 1 star, and 1 line each in the other 5 stars.
These, with considerable uncertainty, show a large star-to-star scatter
in deduced C abundance.  However, the C I lines in stars with
\teff\ $\lesssim 4200$ K may be blended with or completely
dominated by lines from the 
red system of CN, as illustrated in the spectrum of
Arcturus by \citet{hinkle00}. 
It is very likely that this has happened,
as there are two cool stars in our sample,
M71 1-45 and 1-66, with anomalously high deduced C abundances.
The measurements of \citet{bri01b} show
that these two stars have much stronger CN lines than does
M71 star I, a star of similar \teff\, which yielded
a much more reasonable C abundance.  
Unpublished measurements of the G band of CH in these three stars
by JGC also suggest that the very large C abundances we deduce for
M71 1-45 and 1-66 are spurious.

The molecular band data gives a much clearer picture of the pattern
of C abundance variation in M71 as the samples are much larger
and the C abundance can be inferred with considerable precision
from the strength of the CH band.
Both the CH and CN bands clearly show strong star-to-star variations
on the M71 giant branch \citep[][and references therein]{bri01b} and,
more importantly, 
at the level of the main sequence \citep{coh99}.  The entire
set of molecular band data can be explained by a variation in 
C of about a factor of 2, with a much larger anti-correlated variation
in the N abundance.

\subsection{Other clusters}

\begin{figure}
\epsscale{0.7}
\plotone{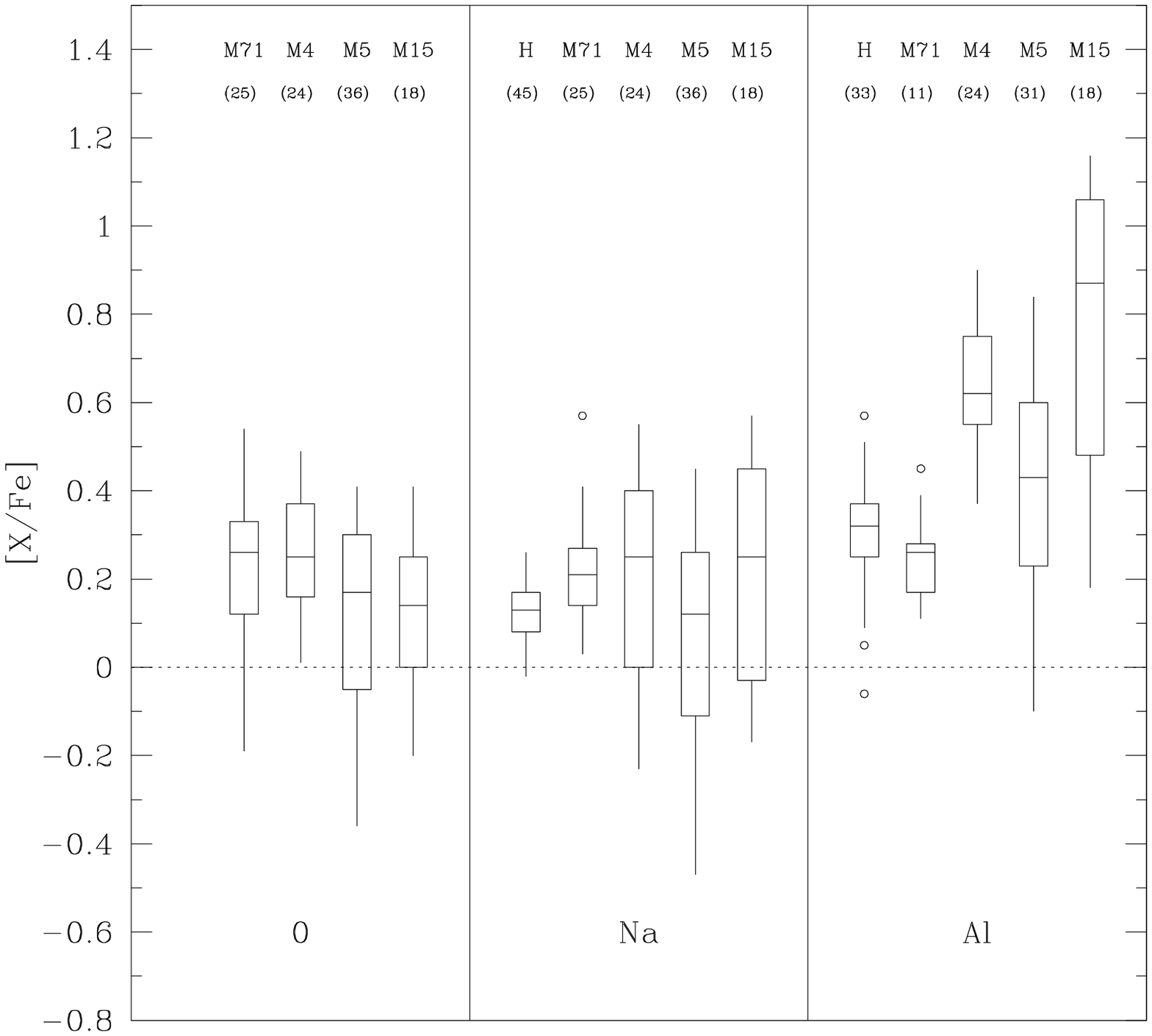}
\figcaption[new_comp1.ps]{A statistical comparison of light elements 
with median abundances from halo dwarfs, M4 ([Fe/H]$\sim$--1.2), M5 ([Fe/H]$\sim$--1.2), 
and M15 from \citet{ful00}, \citet{iva99}, \citet{iva01}, and \citet{sne97}, 
respectively.
The halo dwarfs plotted in the Figures have been selected from the sample
of \citet{ful00} to have [Fe/H] similar to M71 ($-0.6 <$[Fe/H]$<-0.9$).
The number in parenthesis indicates the the number 
of stars analyzed for each element in the corresponding globular cluster.
The layout of the statistical box for each element is as in 
Figure~\ref{summ_fig2}.
\label{new_comp1}}
\end{figure}

\begin{figure}
\epsscale{0.7}
\plotone{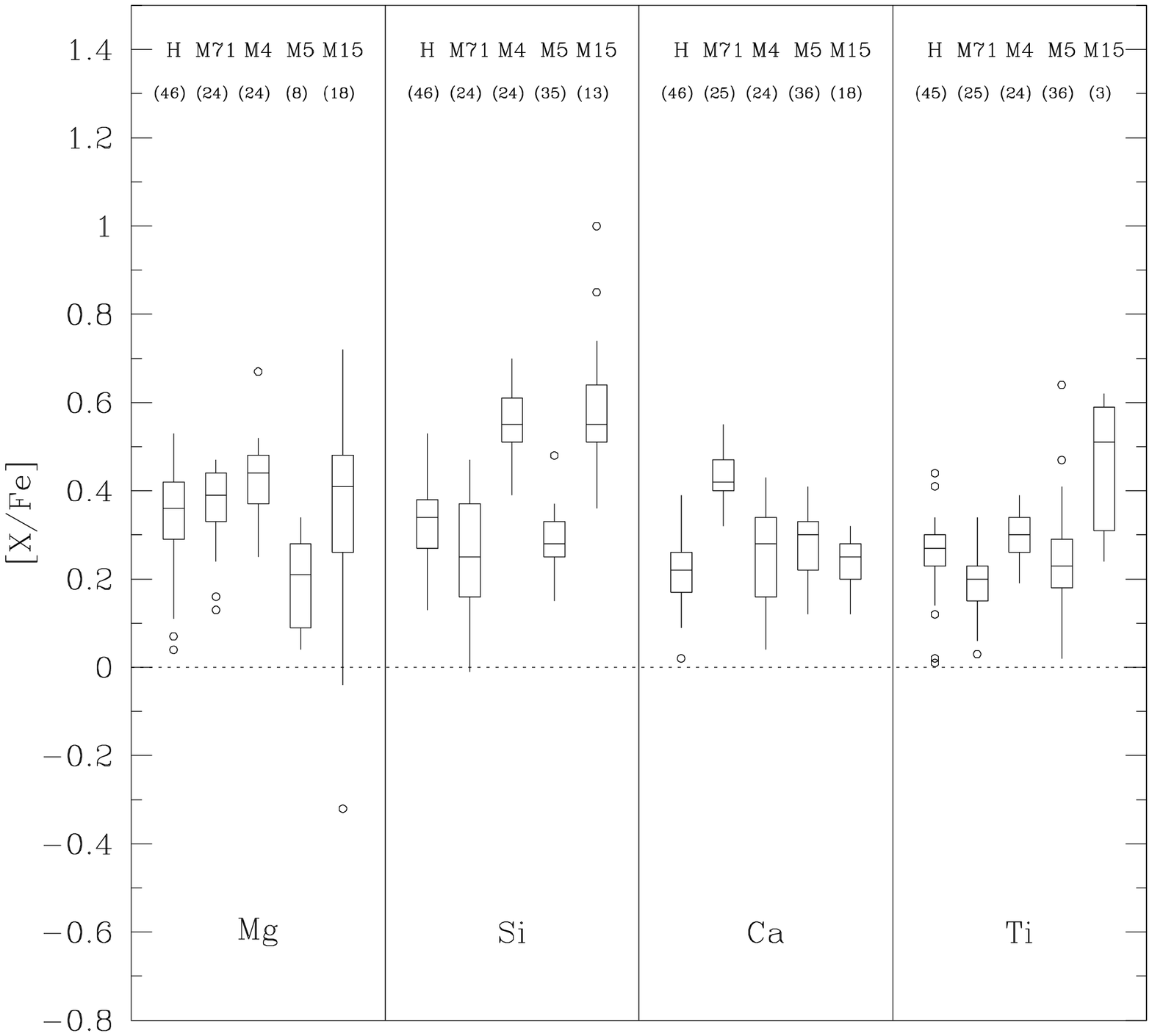}
\figcaption[new_comp2.ps]{Statistical comparison of the abundance of 
$\alpha-$elements in various environments.
Symbols and references as in Figure~\ref{new_comp1}.
\label{new_comp2}}
\end{figure}

\begin{figure}
\epsscale{0.7}
\plotone{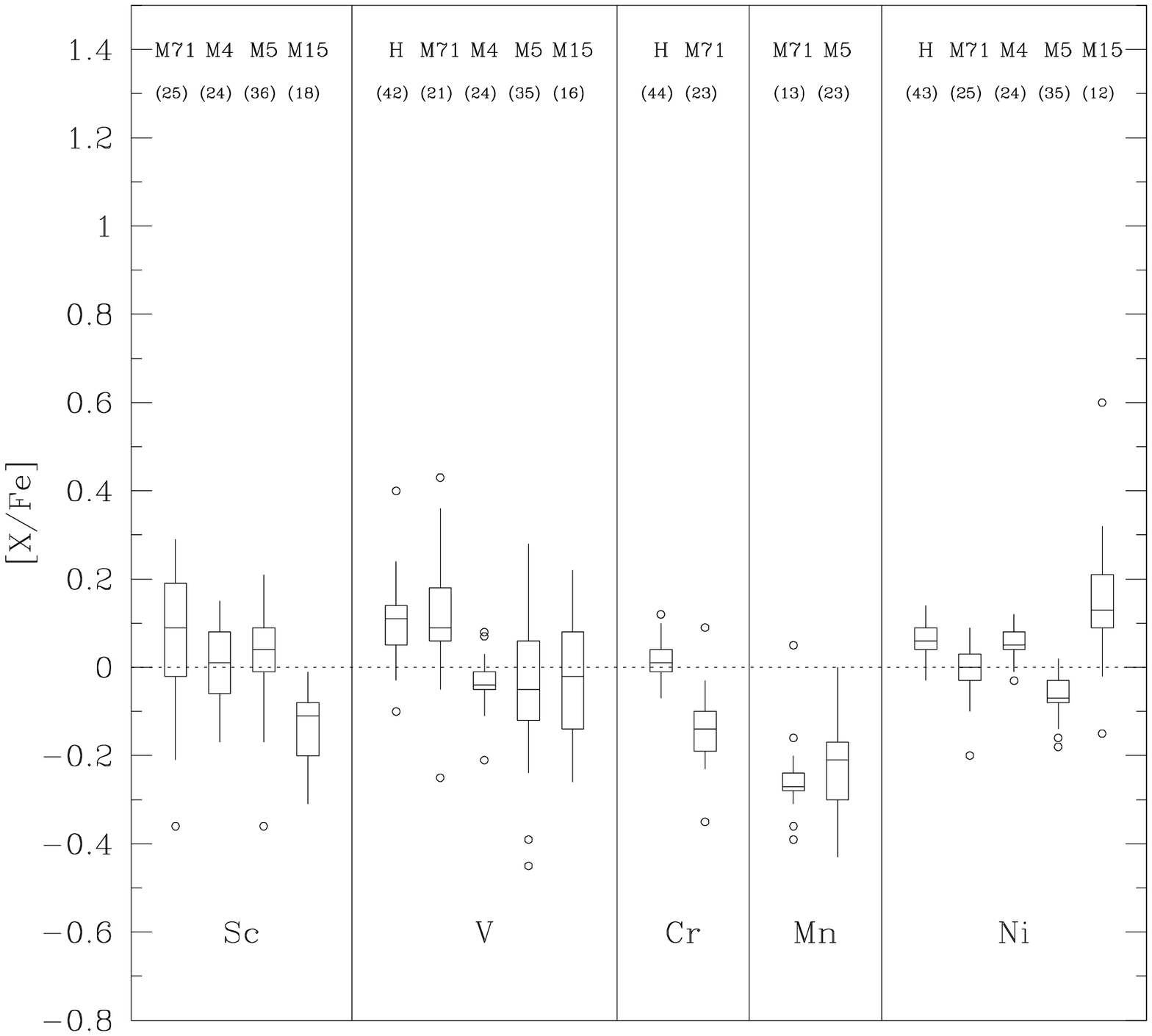}
\figcaption[new_comp3.ps]{Statistical comparison of the abundance of 
iron--peak elements in various environments.
Symbols and references as in Figure~\ref{new_comp1}.
\label{new_comp3}}
\end{figure}

\begin{figure}
\epsscale{0.7}
\plotone{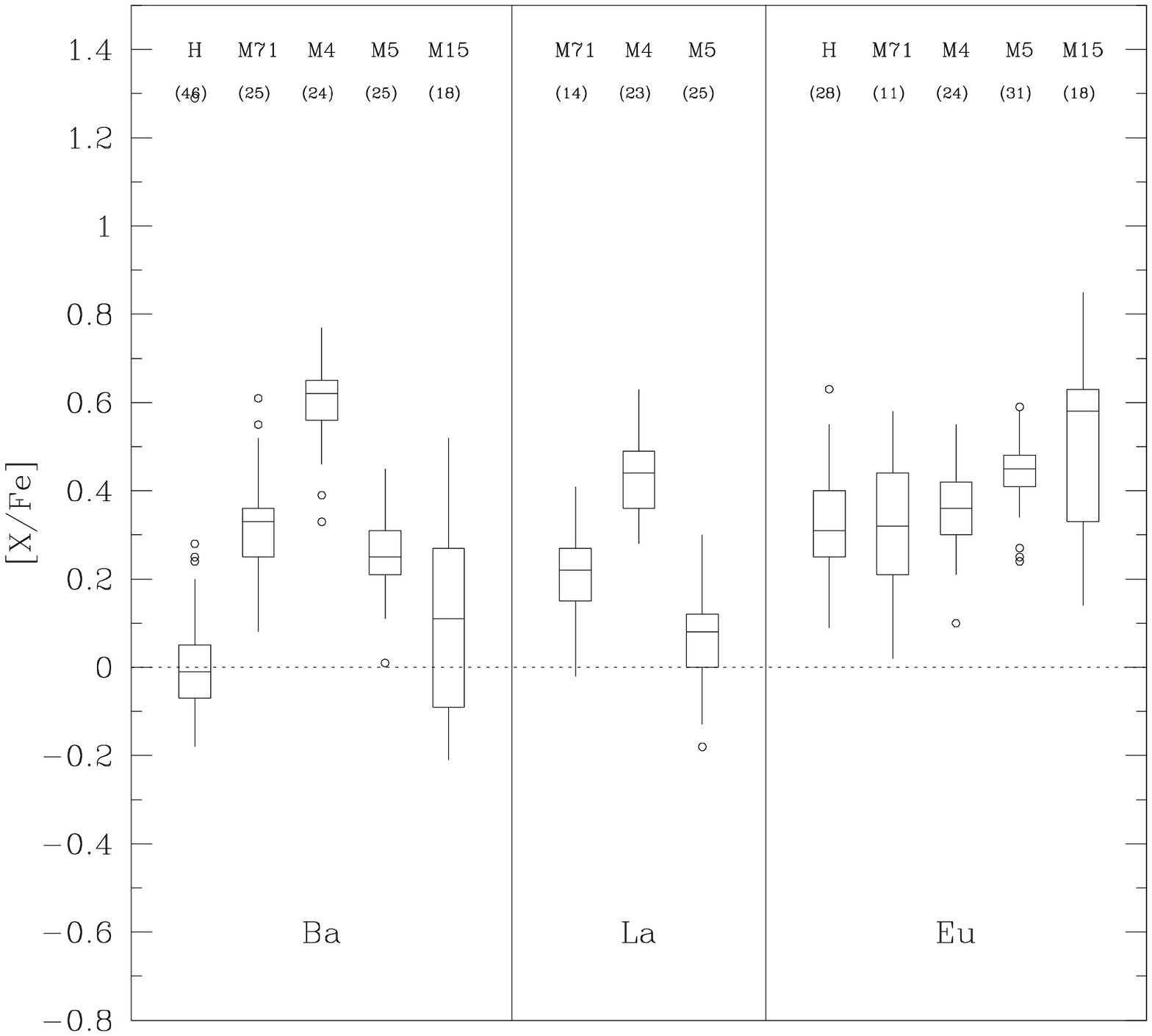}
\figcaption[new_comp4.ps]{Statistical comparison of the abundance of 
neutron capture elements in various environments.
Symbols and references as in Figure~\ref{new_comp1}.
\label{new_comp4}}
\end{figure}

In Figures~\ref{new_comp1} to \ref{new_comp4},  
for each element studied here we provide a comparison to
similar high-resolution abundance analyses of halo dwarfs, of RGB stars in
M4 ([Fe/H]$\sim$--1.2), 
RGB stars in M5 ([Fe/H]$\sim$--1.2), and RGB stars in M15 from \citet{ful00},
\citet{iva99}, \citet{iva01}, and \citet{sne97}, respectively. 
The halo dwarfs plotted in the Figures have been selected from the sample
of \citet{ful00} to have [Fe/H] similar to M71 ($-0.6 <$[Fe/H]$<-0.9$).
The boxes in Figures~\ref{new_comp1} to \ref{new_comp4}
follow the same layout as those in Figure~\ref{summ_fig2}. 
The globular cluster name is indicated above each box ('H' stands for
the halo dwarf sample), and the number in parenthesis below the name 
indicates the number of stars used in the calculation of the respective 
abundance ratio.

The median abundances for each element determined in each of the five
different environments presented in 
Figures~\ref{new_comp1} to \ref{new_comp4} agree to within the
1$\sigma$ uncertainties of each measurement for most of the elements
displayed, and agree to within $\pm1.2\sigma$ for {\it{all}}
the elements shown.  Aluminium is the element showing the largest
trend with metallicity
between these five environments.  This is perhaps
a consequence of not considering non-LTE effects and of some studies
including the 3961 \AA\ doublet, known to be more sensitive to non-LTE
effects, and others not.  Barium is the only other element showing
large variations in its median abundance
among the various environments, although no
consistent trend with metallicity.  We suggest that this too may not
be real, and may be a reflection of the issue of hyperfine structure
corrections in the fairly strong lines of this element.

Overall M71 appears to have very similar
abundance ratios as does M5.  In this set of figures, one can see
some of the well known trends characteristic of halo star abundances
as reviewed by \citet{mcw97}, such as
the gradual increase of [$\alpha$/Fe] as [Fe/H] decreases,
particularly for Si and Ti.

\subsection{Interpretation}

A classical review of post-main sequence stellar evolution can be
found in \citet{ibe83}.  Their description of the consequences
of the first dredge up phase, the only dredge up phase any of 
the stars in our M71 sample may have experienced, 
indicates that a doubling of the surface N$^{14}$ and a 30\% reduction
in the surface C$^{12}$ can be expected, together with a
drop in the ratio of 
C$^{12}$/C$^{13}$ from the solar value of 89 to $\sim$20,
as well as a drop in surface Li and B by several orders of magnitude.
Observations of field stars over a wide range of luminosities
conform fairly well to this picture, see e.g.
\citet{she93,gra00}.

However, the O/Na anti-correlation seen among the bright red giants in 
many globular clusters, including here in the case of M71,
cannot be explained in this picture.  
Several theoretical mechanisms have thus been proposed (e.g., the meridional
mixing of Sweigart \& Mengel 1979, and turbulent diffusion of 
Charbonnel 1994, 1995)
with varying degrees of success.  In addition,
\citet{den90} suggested that the nuclear reaction 
$^{22}$Ne($p,\gamma)^{23}$Na occurs
in regions of the H-burning shell for low mass stars 
where O is converted into N and produces Na$^{23}$ and Al$^{27}$.  
\citet{lan93} combined these ideas to predict the consequences of
such possible synthesis and deep mixing, including for example,
that the surface Mg abundance
should be much less affected than that of Na or Al.  These ideas form the
basis of our current understanding of dredge up in 
low mass metal poor giants,
with more recent calculations given by \citet{den96,cav98,wei00},
among others.

The clear prediction of this suite of calculations is that 
the earliest that deep mixing can
begin is at the location of the bump in the luminosity function
of the RGB which occurs when the H-burning shell crosses a sharp chemical
discontinuity.  \citet{zoc99} have shown
that the luminosity of the RGB bump as
a function of metallicity as determined
from observation agrees well with that predicted by 
the theory of stellar evolution.
\citet{bon01} further suggest that the agreement between the predicted
luminosity function and actual star counts along the RGB in the vicinity of
the bump in a suite of globular clusters is so good that mixing 
(more exactly, mixing of He) cannot have occurred any earlier, otherwise the
evolutionary lifetimes, and hence the observed LF, of such stars 
would have been affected.

Our sample of M71 stars shows a statistically significant correlation
between Al and Na abundances which extends to stars 
as faint as $M_V = +1.8$ mag.  
We also see an anti-correlation between the Na and O 
abundances extending down to near the main sequence turn off.
We see variations in Li (as expected), and may see variations in Zr (not
expected).  Any variations in Mg are smaller than those of Al, Na or O
(as expected).  We know there are large anti-correlated C and N
variations from the work of \citet{coh99}; this too is expected.

The behavior of Li, which is very fragile and easily destroyed, 
is not controversial.
It is, however, the range of luminosity over which the remainder of these
variations are seen which is becoming more and more of a problem
for any scenario which invokes dredge up and mixing.
The Na/O anti-correlation we see in M71 extends to 
the main sequence turn off.  The Al/Na correlation we see in M71
extends to at least as faint as $M_V = +1.8$,
while the RGB bump in a cluster of the
metallicity of M71 is at $M_V = +1.0$.
\citet{coh99} has shown that the C/N anti-correlation
extends to the stars at the main sequence turnoff and even fainter in M71.
\citet{bri01a} have shown that the  C/N abundance range 
seen at the level of the main sequence is comparable to that seen
among the bright red giants of M71 by many previous studies,
the most recent of which is \citet{bri01b}.

The accumulated weight of recent evidence, both in M71 as described above
and in other globular clusters such as 47 Tuc 
\citep[see ][and references therein]{can98}
and NGC 6752 \citep[see][and references therein]{gra01}, suggests that
we are now back in the situation we were during the
late 1980s.  Unless we have missed some important aspect of
stellar evolution with impact on mixing and dredge up, we
must declare the mixing scenario a failure for the specific case
of M71 (and several other globular clusters as well). 
Even the theoreticians in the forefront of this field are beginning
to admit that deep mixing alone is not sufficient \citep{den01,ven01}.
Unless and until some major new concept relevant to this issue appears, 
we must now regard the fundamental origin of the star-to-star
variations we see in M71 as arising outside the stars whose 
spectra we have studied here.

\section{CONCLUSIONS}

We present results of a high dispersion analysis of 23 elements 
to obtain abundance ratios for 25 members of the Galactic 
globular cluster M71.  Our sample of stars includes
19 giant stars (9 of which are less luminous than the RHB), 
3 horizontal branch stars, and 3 stars near 
the main sequence turnoff. Our conclusions are summarized as follows:

\begin{itemize}
\item The iron peak and neutron capture element abundance ratios show no trend 
with \teff, and low scatter around the mean.

\item The $\alpha-$elements Mg, Ca, Si and Ti are overabundant relative to Fe.
The scatter about the mean is small.

\item An anti-correlation between the Na and O abundances for stars in M71
is detected with a statistical
significance in excess of 2$\sigma$, and extends to the 
stars near the main sequence turn off.

\item The [Na/Fe] versus [Al/Fe] correlation is detected with a statistical
significance in excess of 2$\sigma$ in our sample
of M71 stars, and extends at least as faint as $M_V = +1.8$.

\item Both extremely O poor, Na rich stars and extremely O rich, Na poor stars
such as are observed in M5 and in M13 
are not present in our sample of M71 stars.

\item Li is varying  among the subgiants
(as expected), and Zr may be varying among the
subgiants.

\end{itemize}

M71 appears very similar in its element abundance ratios to M5, which
is not surprising as M5 has a metallicity only slightly lower, 
[Fe/H] = $-1.2$ dex \citep{iva01}.   However, the amplitude
of the Na/O and Al/Na relationships appears to be somewhat larger 
in M5 than in M71, and still larger in even more metal poor clusters.

Our detailed abundance analysis of 25 stars in M71 has revealed
abundance variations appearing at such low luminosities
that deep mixing scenarios can no longer reproduce these results.
This problem is made even more acute when we add in
the data of \citet{coh99} and the analysis of
\citet{bri01a} of the CH and CN bands in M71.
We are forced to the firm conclusion that
much, if not all, of the abundance variations seen in 
M71 must have been in place
before the present generation of stars we observe was formed,
or (less likely) are the result of some type (binaries ?) of mass transfer.

In future papers, we will proceed to apply the techniques and analysis
developed here for M71 to other more metal poor 
globular clusters, where the RGB bump is predicted to be somewhat
more luminous and where, judging from the behavior of their
bright RGB stars, we may anticipate finding even larger variations
at low luminosities among the cluster subgiants and main sequence stars.

\acknowledgements
The entire Keck/HIRES user communities owes a huge debt to 
Jerry Nelson, Gerry Smith, Steve Vogt, and many other 
people who have worked to make the Keck Telescope and HIRES  
a reality and to operate and maintain the Keck Observatory. 
We are grateful to the W. M.  Keck Foundation for the vision to fund
the construction of the W. M. Keck Observatory. 
The authors wish to extend special thanks to those of Hawaiian ancestry
on whose sacred mountain we are privileged to be guests. 
Without their generous hospitality, none of the observations presented
herein would have been possible.
We are grateful to the National Science Foundation for partial support under
grant AST-9819614 to JGC.  We thank Jason Prochaska and Andy McWilliam
for providing their tables of hyperfine structure in digital form.


\clearpage

\clearpage
%
%

%
%
\begin{deluxetable}{lcccrcrc}
\tablenum{1}
\tablewidth{0pt}
\tablecaption{Stellar Parameters for the M71 Sample from 
Paper I and Paper II\tablenotemark{a}
\label{tab1}}
\tablehead{
\colhead{ID\tablenotemark{b}} & \colhead{\teff} 
& \colhead{\grav} & \colhead{\mtv}  
& \colhead{${\rm N_{FeI}}$} & \colhead{[Fe/H]$_{\rm FeI}$}
& \colhead{${\rm N_{FeII}}$}& \colhead{[Fe/H]$_{\rm FeII}$} \\
\colhead{}                    & \colhead{(K)}   
& \colhead{}      & \colhead{(km/s)} 
& \colhead{} & \colhead{} & \colhead{} & \colhead{}
}
\startdata 
1--45                   & 3950 & 0.90 & 1.45 & 187 & --0.60$\pm$ 0.03 &   5 & --0.45$\pm$ 0.06\\
I                       & 4150 & 1.00 & 1.38 & 186 & --0.63$\pm$ 0.03 &   6 & --0.66$\pm$ 0.07\\
1--66                   & 4250 & 1.35 & 1.35 & 179 & --0.58$\pm$ 0.03 &   6 & --0.67$\pm$ 0.09\\
1--64                   & 4200 & 1.35 & 1.37 & 187 & --0.61$\pm$ 0.03 &   5 & --0.52$\pm$ 0.09\\
1--56                   & 4525 & 1.60 & 1.26 & 127 & --0.48$\pm$ 0.03 &   2 & --0.68$\pm$ 0.13\\
1--95                   & 4550 & 1.65 & 1.25 & 184 & --0.59$\pm$ 0.03 &   8 & --0.74$\pm$ 0.05\\
1--81                   & 4550 & 1.75 & 1.25 & 180 & --0.56$\pm$ 0.03 &   6 & --0.93$\pm$ 0.05\\
1--1                    & 4700 & 2.05 & 1.20 & 134 & --0.59$\pm$ 0.03 &   5 & --0.76$\pm$ 0.05\\
1--80\tablenotemark{c,d}& 5300 & 2.40 & 1.80 &  71 & --0.64$\pm$ 0.03 &   5 & --0.93$\pm$ 0.05\\
1--87\tablenotemark{c}  & 5300 & 2.40 & 1.80 & 128 & --0.56$\pm$ 0.03 &   9 & --0.84$\pm$ 0.05\\
1--94\tablenotemark{c}  & 5300 & 2.40 & 1.80 &  94 & --0.73$\pm$ 0.03 &   6 & --0.82$\pm$ 0.05\\
1--60                   & 4900 & 2.30 & 1.13 & 119 & --0.74$\pm$ 0.03 &   6 & --0.62$\pm$ 0.05\\
1--59                   & 4600 & 2.30 & 1.23 & 141 & --0.70$\pm$ 0.03 &   5 & --0.58$\pm$ 0.05\\
G53476\_4543            & 4900 & 2.65 & 1.13 & 174 & --0.61$\pm$ 0.03 &   7 & --0.76$\pm$ 0.05\\
2--160                  & 5100 & 2.70 & 1.07 & 145 & --0.49$\pm$ 0.03 &   5 & --0.94$\pm$ 0.08\\
G53447\_4707            & 5175 & 2.75 & 1.04 & 155 & --0.51$\pm$ 0.03 &   7 & --0.83$\pm$ 0.05\\
G53445\_4647            & 5050 & 2.85 & 1.08 & 112 & --0.59$\pm$ 0.03 &   6 & --0.80$\pm$ 0.05\\
G53447\_4703            & 5000 & 3.00 & 1.10 & 125 & --0.71$\pm$ 0.03 &   4 & --0.74$\pm$ 0.05\\
G53425\_4612            & 5150 & 3.15 & 1.05 &  80 & --0.67$\pm$ 0.03 &   2 & --0.87$\pm$ 0.07\\
G53477\_4539            & 5150 & 3.15 & 1.05 & 119 & --0.64$\pm$ 0.03 &   5 & --0.86$\pm$ 0.05\\
G53457\_4709            & 5200 & 3.35 & 1.03 &  93 & --0.72$\pm$ 0.03 &   5 & --0.72$\pm$ 0.12\\
G53391\_4628            & 5100 & 3.35 & 1.07 & 106 & --0.78$\pm$ 0.03 &   5 & --0.76$\pm$ 0.05\\
G53417\_4431            & 5800 & 4.05 & 0.83 &  38 & --0.63$\pm$ 0.04 &   3 & --0.60$\pm$ 0.12\\
G53392\_4624            & 5800 & 4.05 & 0.83 &  36 & --0.76$\pm$ 0.03 &   3 & --0.66$\pm$ 0.08\\
G53414\_4435            & 5900 & 4.15 & 0.80 &  13 & --0.78$\pm$ 0.05 &   2 & --0.58$\pm$ 0.16\\
\enddata
\tablenotetext{a}{$\xi$ and [Fe/H] have been slightly updated from
the values given in Paper I.}
\tablenotetext{b}{Identifications are from \citet{arp71}
or are assigned based on the J2000 coordinates, rh rm rs.s dd dm dd becoming
Grmrss\_dmdd.}
\tablenotetext{c}{RHB star.}
\tablenotetext{d}{Appears to show rotation (Paper I).}
\end{deluxetable}

\clearpage
%
%
\begin{deluxetable}{llrrcccccccccccc}   
\rotate                                 
\tablenum{2}                        
\tablewidth{0pt}                   
\tablecaption{Equivalent Widths (m$\AA$)\tablenotemark{a}\label{tab2}}   
\tablehead{                                                             
\colhead{Ion} & \colhead{$\lambda (\AA)$} & \colhead{$\chi$ (eV)} & \colhead{log($gf$)} & \colhead{1--45 } & \colhead{I    } & \colhead{1--66 } & \colhead{1--64 } & \colhead{1--56 } & \colhead{1--95 } & \colhead{1--81 } & \colhead{1--1  } & \colhead{1--80 } & \colhead{1--87 } & \colhead{1--94 } & \colhead{1--60 } \\                                                       
\colhead{} & \colhead{} & \colhead{} & \colhead{} & \colhead{} & \colhead{} & \colhead{} & \colhead{} & \colhead{} & \colhead{} & \colhead{} & \colhead{} & \colhead{} & \colhead{} & \colhead{} & \colhead{} }          
\startdata    
C I   & 7113.180 &   8.640 &  -0.773 &   35.5                   &   16.1                   &   38.4                   &   ...                    &   ...                    &   30.0                   &   ...                    &   ...                    &   ...                    &   ...                    &   ...                    &   ...                    \\
C I   & 7115.170 &   8.640 &  -0.710 &   31.3                   &   ...                    &   ...                    &   ...                    &   ...                    &   ...                    &   ...                    &   ...                    &   ...                    &   ...                    &   ...                    &   ...                    \\
O I   & 6300.304 &   0.000 &  -9.780 &   76.2                   &   73.8\tablenotemark{b}  &   57.1                   &   71.6                   &   42.4\tablenotemark{b}  &   44.9                   &   44.5                   &   30.5\tablenotemark{b}  &   18.8\tablenotemark{b}  &   23.4                   &   20.1\tablenotemark{b}  &   30.4                   \\
O I   & 6363.776 &   0.020 & -10.300 &   39.3                   &   34.3                   &   25.9                   &   35.7                   &   19.2\tablenotemark{b}  &   19.6                   &   22.3                   &   ...                    &   ...                    &   ...                    &    9.0\tablenotemark{b}  &   14.9\tablenotemark{b}  \\
O I   & 7771.944 &   9.150 &   0.369 &   ...                    &   ...                    &   15.2\tablenotemark{b}  &   21.3                   &   23.2\tablenotemark{b}  &   21.2                   &   19.9                   &   28.7                   &   65.5                   &   52.5                   &   81.0                   &   44.4                   \\
O I   & 7774.166 &   9.150 &   0.223 &   ...                    &   17.2\tablenotemark{b}  &   16.1\tablenotemark{b}  &   18.8\tablenotemark{b}  &   17.5\tablenotemark{b}  &   23.8                   &   17.5\tablenotemark{b}  &   21.4\tablenotemark{b}  &   63.6                   &   50.9                   &   62.3\tablenotemark{b}  &   43.6\tablenotemark{b}  \\
O I   & 7775.388 &   9.150 &   0.001 &   ...                    &   18.6                   &   ...                    &   14.3\tablenotemark{b}  &   18.7\tablenotemark{b}  &   14.7\tablenotemark{b}  &   14.7\tablenotemark{b}  &   14.3\tablenotemark{b}  &   32.3                   &   31.3                   &   50.6\tablenotemark{b}  &   32.5                   \\
Na I  & 5682.633 &   2.100 &  -0.700 &  159.9                   &  130.6                   &  137.0                   &  123.8\tablenotemark{b}  &  121.5                   &  105.0                   &  106.8                   &  106.4                   &   66.1\tablenotemark{b}  &   68.6                   &   63.1                   &   64.4\tablenotemark{b}  \\
Na I  & 5688.193 &   2.100 &  -0.420 &  159.5                   &  134.3                   &  146.3                   &  141.0                   &  131.5                   &  125.5                   &  129.3                   &  125.3                   &   84.4                   &   90.4                   &   83.4                   &   84.9                   \\
Na I  & 6154.225 &   2.100 &  -1.530 &  108.2                   &   60.4                   &   72.3                   &   64.5                   &   52.8                   &   43.0                   &   47.3                   &   48.4                   &    7.0\tablenotemark{b}  &   15.7\tablenotemark{b}  &   13.4\tablenotemark{b}  &   18.4\tablenotemark{b}  \\
Na I  & 6160.747 &   2.100 &  -1.230 &  136.6                   &   91.1\tablenotemark{b}  &  101.3                   &   93.3                   &   89.5                   &   64.3                   &   71.2                   &   68.6                   &   27.6                   &   32.0                   &   17.3                   &   30.6\tablenotemark{b}  \\
Mg I  & 5711.088 &   4.340 &  -1.670 &  137.7                   &  139.4                   &  136.0                   &  126.3                   &  128.1                   &  126.3                   &  124.1                   &  103.9                   &  100.1                   &  102.2                   &   97.9                   &  107.3                   \\
Mg I  & 6318.717 &   5.110 &  -1.970 &   69.0                   &   63.8                   &   55.9                   &   56.7                   &   64.2                   &   53.3                   &   54.9                   &   61.2                   &   ...                    &   37.9                   &   ...                    &   37.4                   \\
Mg I  & 6319.237 &   5.110 &  -2.220 &   55.9                   &   51.0                   &   50.2                   &   53.9                   &   ...                    &   38.7                   &   43.1                   &   ...                    &   ...                    &   26.9                   &   25.3                   &   ...                    \\
Mg I  & 6319.495 &   5.110 &  -2.680 &   ...                    &   ...                    &   ...                    &   ...                    &   ...                    &   ...                    &   ...                    &   ...                    &   ...                    &   ...                    &   ...                    &   ...                    \\
Mg I  & 6965.409 &   5.750 &  -1.870 &   35.7                   &   34.9                   &   41.8                   &   40.8                   &   ...                    &   37.6                   &   42.5                   &   31.3                   &   ...                    &   ...                    &   ...                    &   ...                    \\
Mg I  & 7193.184 &   5.750 &  -1.400 &   68.4                   &   61.4                   &   71.1                   &   63.7                   &   55.5                   &   53.5                   &   52.9                   &   ...                    &   ...                    &   41.0                   &   ...                    &   37.3                   \\
Mg I  & 7387.689 &   5.750 &  -0.870 &   51.3                   &   66.9                   &   62.2                   &   54.4                   &   ...                    &   52.9                   &   61.4                   &   48.8                   &   ...                    &   51.6                   &   41.6                   &   ...                    \\
Mg I  & 7657.603 &   5.110 &  -1.280 &  112.5                   &  112.5                   &  111.6                   &  112.5                   &   97.3                   &   98.0                   &  101.9                   &   91.4                   &   77.4                   &   77.4                   &   ...                    &   ...                    \\
Mg I  & 7722.601 &   5.940 &  -1.800 &   ...                    &   ...                    &   35.6                   &   ...                    &   ...                    &   32.8                   &   46.8                   &   ...                    &   ...                    &   ...                    &   ...                    &   ...                    \\
Mg I  & 7930.810 &   5.940 &  -1.200 &   53.6                   &   60.2                   &   67.9                   &   56.5                   &   ...                    &   62.2                   &   52.6                   &   45.0                   &   ...                    &   ...                    &   ...                    &   ...                    \\
Al I  & 5557.059 &   3.140 &  -2.100 &   ...                    &   ...                    &   29.2                   &   22.0                   &   ...                    &   14.1                   &   18.8                   &   ...                    &   ...                    &   ...                    &   ...                    &   ...                    
\enddata  
\tablenotetext{a}{Table available electronically.}  
\tablenotetext{b}{Line identified by hand. All other lines are   
identified automatically.}                                      
\tablenotetext{c}{Fe I line used in the $\lambda D-W_{\lambda} $ fit.}
\end{deluxetable}                                                    

\clearpage
%
%
\begin{deluxetable}{lcc}
\tablenum{3}
\tablewidth{0pt}
\tablecaption{Correction Factors for Inverted Solar $gf$ Values 
\label{tab3}}
\tablehead{\colhead{Ion} & \colhead{\# Common Lines} & \colhead{Correction} \\
\colhead{} & \colhead{Between NIST \& Solar} & 
\colhead{Factor\tablenotemark{a}} \\
\colhead{} & \colhead{} & \colhead{(dex)} }
\startdata
Mg I  &  4 & +0.10 \\
Al I  &  6 & +0.21 \\
Ca I  & 12 & +0.33 \\
Ti I  & 30 & +0.05 \\
Cr I  & 11 & +0.05 \\
Ni I  & 33 & +0.05 \\
\enddata
\tablenotetext{a}{$gf$(used) = $gf$(Thevenin) + Correction Factor}
\end{deluxetable}

\clearpage
%
%
\begin{deluxetable}{lcccc}
\tablenum{4}
\tablewidth{0pt}
\tablecaption{Solar Abundance Ratios [X/Fe] 
\label{tab4}}
\tablehead{\colhead{Ion} & \colhead{\# lines} & 
\colhead{[X/Fe]\tablenotemark{a}} & 
\colhead{$\sigma$\tablenotemark{a}}
& \colhead{$\Delta$[us-meteoric]\tablenotemark{b}} \\
\colhead{} & \colhead{} & \colhead{(dex)} & \colhead{(dex)} & \colhead{(dex)}}
\startdata
Li I\tablenotemark{c} &  1 &  0.94  & ...  & --0.22\tablenotemark{d} \\
C I   &  4 &  +1.08 & 0.04 &  +0.03 \\
O I   &  5 &  +1.53 & 0.10 &  +0.11 \\
Na I  &  4 & --1.30 & 0.09 & --0.10 \\
Mg I  & 10 &  +0.03 & 0.24 & --0.05 \\
Al I  &  6 & --1.20 & 0.15 & --0.17 \\
Si I  & 20 &  +0.14 & 0.12 &  +0.09 \\
K I   &  1 & --2.28 & ...  &  +0.10 \\
Ca I  & 15 & --1.56 & 0.14 & --0.39 \\
Sc II &  7 & --4.26 & 0.12 &  +0.16 \\
Ti I  & 40 & --2.48 & 0.15 &  +0.09 \\
V I   & 13 & --3.55 & 0.14 & --0.06 \\
Cr I  & 12 & --1.72 & 0.16 &  +0.10 \\
Mn I  &  4 & --2.09 & 0.12 & --0.12 \\
Co I  &  7 & --2.60 & 0.08 &   0.00 \\
Ni I  & 43 & --1.19 & 0.18 &  +0.07 \\
Cu I  &  1 & --3.44 & ...  & --0.20 \\
Zn I  &  1 & --2.94 & ...  & --0.09 \\
Y II  &  1 & --4.96 & ...  &  +0.33 \\
Zr I  &  4 & --4.52 & 0.16 &  +0.38 \\
Ba II &  3 & --5.29 & 0.08 &  +0.01 \\
La II &  3 & --6.27 & 0.07 &  +0.04 \\
Eu II &  1 & --6.96 & ...  &  +0.01 \\
\enddata
\tablenotetext{a}{Mean and 1$\sigma$ rms deviation about the mean
for the abundance in the Sun of the lines 
of a particular ion using
our adopted atomic line parameters.}
\tablenotetext{b}{Meteoric solar abundances from \citet{and89}.}
\tablenotetext{c}{log $\epsilon$(Li)}
\tablenotetext{d}{The photospheric Solar Li abundance
from \citet{and89} is used.}
\end{deluxetable}

%
%
\clearpage

\begin{deluxetable}{lrrrrrrrr}  
\tablenum{5a}                    
\tablewidth{0pt}                
\tablecaption{Abundance Ratios/Li-Na.\label{tab5a}} 
\tablehead{                     
\colhead{Star} &                
\colhead{${\rm N_{Li}}$} & \colhead{[Li/Fe]} &
\colhead{${\rm N_{C }}$} & \colhead{[C /Fe]} &
\colhead{${\rm N_{O }}$} & \colhead{[O /Fe]} &
\colhead{${\rm N_{Na}}$} & \colhead{[Na/Fe]} }
\startdata
1--45        &   0 &        ...      &   2 &   2.10$\pm$0.12 &   2 & --0.19$\pm$0.07 &   4 &   0.57$\pm$0.08\\
I            &   0 &        ...      &   1 &   1.20$\pm$0.10 &   4 &   0.20$\pm$0.18 &   4 &   0.22$\pm$0.08\\
1--66        &   1 &$<$ --0.61       &   1 &   1.93$\pm$0.11 &   4 &   0.03$\pm$0.08 &   4 &   0.41$\pm$0.07\\
1--64        &   1 &$<$ --0.79       &   0 &        ...      &   5 &   0.14$\pm$0.07 &   4 &   0.25$\pm$0.05\\
1--56        &   0 &        ...      &   0 &        ...      &   5 &   0.01$\pm$0.10 &   4 &   0.34$\pm$0.08\\
1--95        &   1 & $<$--0.39       &   1 &   1.32$\pm$0.09 &   5 &   0.09$\pm$0.11 &   4 &   0.23$\pm$0.08\\
1--81        &   0 &        ...      &   0 &        ...      &   5 &   0.27$\pm$0.14 &   4 &   0.27$\pm$0.07\\
1--1         &   1 &  $<$ 0.36       &   0 &        ...      &   4 &   0.10$\pm$0.07 &   4 &   0.39$\pm$0.06\\
1--80        &   1 & $<$--0.69       &   0 &        ...      &   4 &   0.53$\pm$0.19 &   4 &   0.03$\pm$0.12\\
1--87        &   0 &        ...      &   0 &        ...      &   4 &   0.34$\pm$0.07 &   4 &   0.10$\pm$0.05\\
1--94        &   1 & $<$--0.69       &   0 &        ...      &   5 &   0.53$\pm$0.21 &   4 &   0.12$\pm$0.08\\
1--60        &   0 &        ...      &   0 &        ...      &   5 &   0.33$\pm$0.12 &   4 &   0.04$\pm$0.05\\
1--59        &   0 &        ...      &   0 &        ...      &   5 &   0.15$\pm$0.12 &   4 &   0.11$\pm$0.05\\
G53476\_4543 &   0 &        ...      &   0 &        ...      &   5 &   0.26$\pm$0.07 &   4 &   0.15$\pm$0.05\\
2--160       &   1 & $<$  0.61       &   1 &   1.05$\pm$0.11 &   4 &   0.06$\pm$0.14 &   3 &   0.34$\pm$0.08\\
G53447\_4707 &   1 &   1.01$\pm$0.10 &   0 &        ...      &   4 &   0.33$\pm$0.07 &   4 &   0.14$\pm$0.05\\
G53445\_4647 &   1 &   0.99$\pm$0.10 &   1 &   0.74$\pm$0.09 &   4 &   0.31$\pm$0.09 &   4 &   0.16$\pm$0.05\\
G53447\_4703 &   0 &        ...      &   0 &        ...      &   3 &   0.11$\pm$0.10 &   4 &   0.37$\pm$0.05\\
G53425\_4612 &   1 &   1.29$\pm$0.10 &   0 &        ...      &   3 &   0.34$\pm$0.12 &   4 &   0.21$\pm$0.09\\
G53477\_4539 &   0 &        ...      &   0 &        ...      &   4 &   0.33$\pm$0.07 &   4 &   0.06$\pm$0.07\\
G53457\_4709 &   0 &        ...      &   0 &        ...      &   3 &   0.14$\pm$0.11 &   4 &   0.10$\pm$0.06\\
G53391\_4628 &   0 &        ...      &   0 &        ...      &   3 &   0.32$\pm$0.13 &   4 &   0.27$\pm$0.05\\
G53417\_4431 &   0 &        ...      &   0 &        ...      &   3 &   0.12$\pm$0.11 &   4 &   0.28$\pm$0.07\\
G53392\_4624 &   0 &        ...      &   0 &        ...      &   3 &   0.37$\pm$0.09 &   3 &   0.19$\pm$0.10\\
G53414\_4435 &   0 &        ...      &   0 &        ...      &   3 &   0.54$\pm$0.18 &   3 &   0.14$\pm$0.14
\enddata                                          
\end{deluxetable}                                 
\clearpage                      
 
\begin{deluxetable}{lrrrrrrrrrr}
\rotate                         
\tablenum{5b}                    
\tablewidth{0pt}                
\tablecaption{Abundance Ratios/Mg-Ca.\label{tab5b}} 
\tablehead{                     
\colhead{Star} &                
\colhead{${\rm N_{Mg}}$} & \colhead{[Mg/Fe]} &
\colhead{${\rm N_{Al}}$} & \colhead{[Al/Fe]} &
\colhead{${\rm N_{Si}}$} & \colhead{[Si/Fe]} &
\colhead{${\rm N_{K }}$} & \colhead{[K /Fe]} &
\colhead{${\rm N_{Ca}}$} & \colhead{[Ca/Fe]} }
\startdata
1--45        &   8 &   0.43$\pm$0.09 &   3 &   0.45$\pm$0.09 &  12 &   0.04$\pm$0.07 &   1 & --0.71$\pm$0.09 &  15 &   0.32$\pm$0.07\\
I            &   8 &   0.46$\pm$0.06 &   3 &   0.28$\pm$0.11 &  14 &   0.18$\pm$0.08 &   1 & --0.12$\pm$0.09 &  13 &   0.42$\pm$0.06\\
1--66        &   9 &   0.47$\pm$0.09 &   3 &   0.26$\pm$0.07 &  17 &   0.22$\pm$0.10 &   1 &   0.03$\pm$0.09 &  14 &   0.39$\pm$0.06\\
1--64        &   8 &   0.41$\pm$0.09 &   4 &   0.19$\pm$0.07 &  16 &   0.15$\pm$0.10 &   1 & --0.82$\pm$0.09 &  15 &   0.39$\pm$0.06\\
1--56        &   4 &   0.33$\pm$0.06 &   0 &        ...      &  10 &   0.34$\pm$0.14 &   0 &        ...      &  10 &   0.43$\pm$0.07\\
1--95        &   9 &   0.44$\pm$0.09 &   3 &   0.13$\pm$0.07 &  17 &   0.36$\pm$0.06 &   1 & --0.17$\pm$0.09 &  13 &   0.42$\pm$0.06\\
1--81        &   9 &   0.46$\pm$0.09 &   4 &   0.27$\pm$0.07 &  16 &   0.42$\pm$0.06 &   1 &   0.03$\pm$0.09 &  14 &   0.47$\pm$0.06\\
1--1         &   6 &   0.33$\pm$0.11 &   1 &   0.27$\pm$0.07 &  14 &   0.27$\pm$0.06 &   1 & --0.06$\pm$0.09 &  12 &   0.55$\pm$0.06\\
1--80        &   2 &   0.35$\pm$0.06 &   0 &        ...      &  10 &   0.47$\pm$0.06 &   0 &        ...      &  10 &   0.52$\pm$0.13\\
1--87        &   6 &   0.35$\pm$0.06 &   0 &        ...      &  16 &   0.45$\pm$0.06 &   0 &        ...      &  12 &   0.47$\pm$0.08\\
1--94        &   3 &   0.39$\pm$0.13 &   2 &   0.11$\pm$0.07 &  13 &   0.41$\pm$0.06 &   1 & --0.02$\pm$0.09 &  10 &   0.42$\pm$0.06\\
1--60        &   3 &   0.47$\pm$0.06 &   0 &        ...      &  12 &   0.20$\pm$0.06 &   0 &        ...      &  11 &   0.47$\pm$0.06\\
1--59        &   4 &   0.43$\pm$0.06 &   0 &        ...      &  16 &   0.16$\pm$0.06 &   0 &        ...      &  12 &   0.44$\pm$0.06\\
G53476\_4543 &   5 &   0.36$\pm$0.06 &   0 &        ...      &  18 &   0.27$\pm$0.06 &   0 &        ...      &  15 &   0.41$\pm$0.06\\
2--160       &   7 &   0.33$\pm$0.08 &   2 &   0.17$\pm$0.07 &  17 &   0.45$\pm$0.09 &   0 &        ...      &  13 &   0.48$\pm$0.06\\
G53447\_4707 &   7 &   0.32$\pm$0.06 &   2 &   0.21$\pm$0.12 &  17 &   0.38$\pm$0.06 &   1 & --0.19$\pm$0.09 &  12 &   0.48$\pm$0.06\\
G53445\_4647 &   4 &   0.24$\pm$0.06 &   1 &   0.39$\pm$0.07 &  13 &   0.28$\pm$0.06 &   0 &        ...      &  12 &   0.44$\pm$0.07\\
G53447\_4703 &   3 &   0.44$\pm$0.06 &   0 &        ...      &  16 &   0.23$\pm$0.06 &   1 &   0.19$\pm$0.09 &  13 &   0.36$\pm$0.06\\
G53425\_4612 &   5 &   0.39$\pm$0.16 &   0 &        ...      &   4 &   0.14$\pm$0.10 &   1 &   0.00$\pm$0.09 &   9 &   0.47$\pm$0.06\\
G53477\_4539 &   6 &   0.43$\pm$0.06 &   0 &        ...      &  17 &   0.30$\pm$0.06 &   0 &        ...      &  11 &   0.38$\pm$0.06\\
G53457\_4709 &   4 &   0.45$\pm$0.10 &   0 &        ...      &  11 &   0.11$\pm$0.13 &   0 &        ...      &  11 &   0.50$\pm$0.07\\
G53391\_4628 &   2 &   0.13$\pm$0.06 &   0 &        ...      &  15 &   0.17$\pm$0.06 &   0 &        ...      &  10 &   0.39$\pm$0.06\\
G53417\_4431 &   1 &   0.16$\pm$0.06 &   0 &        ...      &   5 &   0.02$\pm$0.13 &   0 &        ...      &  11 &   0.40$\pm$0.06\\
G53392\_4624 &   1 &   0.39$\pm$0.06 &   0 &        ...      &   4 & --0.01$\pm$0.12 &   0 &        ...      &   8 &   0.37$\pm$0.07\\
G53414\_4435 &   0 &        ...      &   0 &        ...      &   0 &        ...      &   0 &        ...      &   8 &   0.42$\pm$0.10
\enddata                                          
\end{deluxetable}                                 
\clearpage                      
 
\begin{deluxetable}{lrrrrrrrr}  
\tablenum{5c}                    
\tablewidth{0pt}                
\tablecaption{Abundance Ratios/Sc-Cr.\label{tab5c}} 
\tablehead{                     
\colhead{Star} &                
\colhead{${\rm N_{Sc}}$} & \colhead{[Sc/Fe]} &
\colhead{${\rm N_{Ti}}$} & \colhead{[Ti/Fe]} &
\colhead{${\rm N_{V }}$} & \colhead{[V /Fe]} &
\colhead{${\rm N_{Cr}}$} & \colhead{[Cr/Fe]} }
\startdata
1--45        &   6 &   0.23$\pm$0.08 &  36 &   0.34$\pm$0.06 &  14 &   0.43$\pm$0.12 &  10 & --0.03$\pm$0.08\\
I            &   7 &   0.29$\pm$0.21 &  37 &   0.23$\pm$0.05 &  14 &   0.18$\pm$0.15 &   9 & --0.10$\pm$0.08\\
1--66        &   6 &   0.21$\pm$0.07 &  38 &   0.23$\pm$0.05 &  13 &   0.03$\pm$0.07 &   9 & --0.11$\pm$0.07\\
1--64        &   7 &   0.19$\pm$0.12 &  37 &   0.25$\pm$0.05 &  14 &   0.19$\pm$0.15 &   8 & --0.10$\pm$0.07\\
1--56        &   4 & --0.04$\pm$0.12 &  21 &   0.15$\pm$0.07 &   9 &   0.08$\pm$0.14 &   3 & --0.21$\pm$0.13\\
1--95        &   6 &   0.15$\pm$0.06 &  36 &   0.06$\pm$0.04 &  14 &   0.02$\pm$0.08 &   9 & --0.23$\pm$0.07\\
1--81        &   7 &   0.07$\pm$0.07 &  34 &   0.34$\pm$0.05 &  14 &   0.36$\pm$0.12 &   9 & --0.07$\pm$0.07\\
1--1         &   7 &   0.20$\pm$0.19 &  25 &   0.22$\pm$0.05 &  14 &   0.19$\pm$0.07 &   5 & --0.14$\pm$0.07\\
1--80        &   4 & --0.36$\pm$0.15 &   8 &   0.17$\pm$0.04 &   1 & --0.05$\pm$0.07 &   1 &   0.09$\pm$0.07\\
1--87        &   7 & --0.17$\pm$0.06 &  16 &   0.13$\pm$0.04 &  10 &   0.20$\pm$0.07 &   4 & --0.21$\pm$0.07\\
1--94        &   6 & --0.02$\pm$0.06 &   5 &   0.07$\pm$0.04 &   4 &   0.03$\pm$0.07 &   3 & --0.35$\pm$0.10\\
1--60        &   5 &   0.20$\pm$0.07 &  10 &   0.03$\pm$0.06 &   3 & --0.25$\pm$0.18 &   3 & --0.19$\pm$0.14\\
1--59        &   5 &   0.26$\pm$0.06 &  23 &   0.13$\pm$0.04 &  11 &   0.06$\pm$0.07 &   7 & --0.11$\pm$0.07\\
G53476\_4543 &   6 &   0.14$\pm$0.06 &  32 &   0.20$\pm$0.04 &  11 &   0.16$\pm$0.07 &   9 & --0.15$\pm$0.07\\
2--160       &   6 &   0.03$\pm$0.06 &  24 &   0.27$\pm$0.05 &  12 &   0.17$\pm$0.07 &   7 & --0.08$\pm$0.08\\
G53447\_4707 &   7 & --0.07$\pm$0.07 &  21 &   0.16$\pm$0.04 &   9 &   0.09$\pm$0.07 &   8 & --0.18$\pm$0.07\\
G53445\_4647 &   7 & --0.02$\pm$0.06 &  15 &   0.20$\pm$0.04 &   8 &   0.18$\pm$0.08 &   6 & --0.10$\pm$0.07\\
G53447\_4703 &   7 &   0.29$\pm$0.21 &  17 &   0.21$\pm$0.04 &   6 &   0.05$\pm$0.08 &   6 & --0.06$\pm$0.07\\
G53425\_4612 &   5 &   0.06$\pm$0.13 &   6 &   0.23$\pm$0.05 &   7 &   0.08$\pm$0.07 &   3 & --0.15$\pm$0.12\\
G53477\_4539 &   5 &   0.02$\pm$0.06 &  13 &   0.19$\pm$0.04 &   7 &   0.15$\pm$0.09 &   3 & --0.23$\pm$0.13\\
G53457\_4709 &   6 & --0.07$\pm$0.06 &   8 &   0.23$\pm$0.04 &   3 &   0.09$\pm$0.07 &   3 & --0.12$\pm$0.08\\
G53391\_4628 &   6 &   0.10$\pm$0.06 &  15 &   0.30$\pm$0.04 &   0 &        ...      &   3 & --0.18$\pm$0.07\\
G53417\_4431 &   3 & --0.21$\pm$0.06 &   6 &   0.27$\pm$0.07 &   0 &        ...      &   1 & --0.17$\pm$0.07\\
G53392\_4624 &   3 &   0.09$\pm$0.11 &   6 &   0.06$\pm$0.09 &   0 &        ...      &   0 &        ...     \\
G53414\_4435 &   1 &   0.15$\pm$0.07 &   4 &   0.15$\pm$0.16 &   0 &        ...      &   0 &        ...     
\enddata                                          
\end{deluxetable}                                 
\clearpage                      
 
\begin{deluxetable}{lrrrrrrrrrr}
\rotate                         
\tablenum{5d}                    
\tablewidth{0pt}                
\tablecaption{Abundance Ratios/Mn-Zn.\label{tab5d}} 
\tablehead{                     
\colhead{Star} &                
\colhead{${\rm N_{Mn}}$} & \colhead{[Mn/Fe]} &
\colhead{${\rm N_{Co}}$} & \colhead{[Co/Fe]} &
\colhead{${\rm N_{Ni}}$} & \colhead{[Ni/Fe]} &
\colhead{${\rm N_{Cu}}$} & \colhead{[Cu/Fe]} &
\colhead{${\rm N_{Zn}}$} & \colhead{[Zn/Fe]} }
\startdata
1--45        &   3 & --0.28$\pm$0.22 &   7 &   0.09$\pm$0.08 &  37 &   0.09$\pm$0.05 &   1 & --0.12$\pm$0.09 &   0 &        ...     \\
I            &   4 & --0.24$\pm$0.06 &   7 &   0.02$\pm$0.07 &  35 &   0.05$\pm$0.05 &   1 &   0.05$\pm$0.09 &   0 &        ...     \\
1--66        &   4 & --0.25$\pm$0.06 &   7 &   0.01$\pm$0.05 &  35 &   0.06$\pm$0.05 &   1 &   0.14$\pm$0.09 &   1 &   0.74$\pm$0.12\\
1--64        &   4 & --0.39$\pm$0.06 &   8 &   0.02$\pm$0.05 &  38 &   0.08$\pm$0.05 &   1 & --0.22$\pm$0.09 &   1 &   0.64$\pm$0.12\\
1--56        &   0 &        ...      &   3 &   0.00$\pm$0.05 &  29 &   0.04$\pm$0.06 &   0 &        ...      &   0 &        ...     \\
1--95        &   4 & --0.36$\pm$0.06 &   7 & --0.02$\pm$0.06 &  37 &   0.03$\pm$0.05 &   1 &   0.12$\pm$0.09 &   0 &        ...     \\
1--81        &   4 & --0.16$\pm$0.06 &   7 &   0.04$\pm$0.05 &  36 &   0.00$\pm$0.05 &   1 &   0.12$\pm$0.09 &   0 &        ...     \\
1--1         &   4 & --0.27$\pm$0.06 &   6 &   0.04$\pm$0.06 &  29 &   0.02$\pm$0.06 &   1 &   0.07$\pm$0.09 &   0 &        ...     \\
1--80        &   0 &        ...      &   0 &        ...      &  17 & --0.07$\pm$0.06 &   1 &   0.38$\pm$0.09 &   0 &        ...     \\
1--87        &   4 & --0.24$\pm$0.09 &   4 &   0.13$\pm$0.08 &  27 & --0.07$\pm$0.05 &   1 &   0.16$\pm$0.09 &   1 &   0.46$\pm$0.09\\
1--94        &   3 & --0.31$\pm$0.06 &   0 &        ...      &  16 &   0.00$\pm$0.06 &   1 &   0.09$\pm$0.09 &   1 &   0.44$\pm$0.09\\
1--60        &   0 &        ...      &   2 &   0.04$\pm$0.12 &  19 &   0.00$\pm$0.05 &   1 & --0.30$\pm$0.09 &   0 &        ...     \\
1--59        &   0 &        ...      &   3 &   0.03$\pm$0.07 &  28 &   0.07$\pm$0.05 &   1 &   0.04$\pm$0.09 &   0 &        ...     \\
G53476\_4543 &   0 &        ...      &   3 &   0.05$\pm$0.05 &  36 & --0.02$\pm$0.05 &   1 &   0.07$\pm$0.09 &   0 &        ...     \\
2--160       &   3 & --0.28$\pm$0.06 &   7 &   0.12$\pm$0.05 &  28 & --0.03$\pm$0.06 &   1 &   0.14$\pm$0.09 &   1 &   0.42$\pm$0.11\\
G53447\_4707 &   3 & --0.27$\pm$0.06 &   4 & --0.05$\pm$0.08 &  34 & --0.05$\pm$0.05 &   1 &   0.00$\pm$0.09 &   1 &   0.44$\pm$0.09\\
G53445\_4647 &   0 &        ...      &   2 &   0.00$\pm$0.17 &  23 &   0.00$\pm$0.07 &   1 &   0.19$\pm$0.09 &   1 &   0.34$\pm$0.09\\
G53447\_4703 &   2 &   0.05$\pm$0.12 &   1 &   0.04$\pm$0.05 &  25 &   0.05$\pm$0.05 &   1 &   0.15$\pm$0.09 &   0 &        ...     \\
G53425\_4612 &   2 & --0.20$\pm$0.10 &   0 &        ...      &  13 &   0.02$\pm$0.10 &   1 &   0.08$\pm$0.09 &   0 &        ...     \\
G53477\_4539 &   0 &        ...      &   0 &        ...      &  24 & --0.03$\pm$0.06 &   1 &   0.09$\pm$0.09 &   0 &        ...     \\
G53457\_4709 &   0 &        ...      &   0 &        ...      &  15 & --0.01$\pm$0.07 &   1 & --0.01$\pm$0.09 &   1 &   0.20$\pm$0.14\\
G53391\_4628 &   0 &        ...      &   2 &   0.14$\pm$0.15 &  20 & --0.03$\pm$0.07 &   1 &   0.15$\pm$0.09 &   0 &        ...     \\
G53417\_4431 &   0 &        ...      &   0 &        ...      &   7 & --0.10$\pm$0.14 &   0 &        ...      &   0 &        ...     \\
G53392\_4624 &   0 &        ...      &   0 &        ...      &   5 & --0.20$\pm$0.15 &   0 &        ...      &   0 &        ...     \\
G53414\_4435 &   0 &        ...      &   0 &        ...      &   2 & --0.01$\pm$0.06 &   0 &        ...      &   0 &        ...     
\enddata                                          
\end{deluxetable}                                 
\clearpage                      
 
\begin{deluxetable}{lrrrrrrrrrr}
\rotate                         
\tablenum{5e}                    
\tablewidth{0pt}                
\tablecaption{Abundance Ratios/Y-Eu.\label{tab5e}} 
\tablehead{                     
\colhead{Star} &                
\colhead{${\rm N_{Y }}$} & \colhead{[Y /Fe]} &
\colhead{${\rm N_{Zr}}$} & \colhead{[Zr/Fe]} &
\colhead{${\rm N_{Ba}}$} & \colhead{[Ba/Fe]} &
\colhead{${\rm N_{La}}$} & \colhead{[La/Fe]} &
\colhead{${\rm N_{Eu}}$} & \colhead{[Eu/Fe]} }
\startdata
1--45        &   0 &        ...      &   3 &   0.09$\pm$0.27 &   3 &   0.25$\pm$0.14 &   2 &   0.22$\pm$0.08 &   1 &   0.58$\pm$0.09\\
I            &   1 & --0.16$\pm$0.09 &   4 & --0.24$\pm$0.10 &   3 &   0.35$\pm$0.08 &   1 &   0.15$\pm$0.08 &   1 &   0.46$\pm$0.09\\
1--66        &   0 &        ...      &   4 & --0.30$\pm$0.13 &   3 &   0.33$\pm$0.09 &   3 &   0.20$\pm$0.08 &   1 &   0.38$\pm$0.09\\
1--64        &   0 &        ...      &   3 & --0.23$\pm$0.08 &   2 &   0.31$\pm$0.08 &   2 &   0.32$\pm$0.12 &   1 &   0.49$\pm$0.09\\
1--56        &   0 &        ...      &   1 & --0.57$\pm$0.08 &   3 &   0.52$\pm$0.20 &   2 &   0.27$\pm$0.24 &   1 &   0.02$\pm$0.09\\
1--95        &   1 & --0.26$\pm$0.09 &   3 & --0.38$\pm$0.08 &   3 &   0.49$\pm$0.09 &   2 &   0.09$\pm$0.08 &   1 &   0.32$\pm$0.09\\
1--81        &   0 &        ...      &   2 &   0.00$\pm$0.08 &   3 &   0.34$\pm$0.12 &   2 &   0.27$\pm$0.12 &   1 &   0.26$\pm$0.09\\
1--1         &   0 &        ...      &   0 &        ...      &   2 &   0.18$\pm$0.09 &   1 &   0.21$\pm$0.08 &   0 &        ...     \\
1--80        &   0 &        ...      &   0 &        ...      &   3 &   0.38$\pm$0.37 &   0 &        ...      &   1 &   0.15$\pm$0.09\\
1--87        &   0 &        ...      &   1 &   0.20$\pm$0.08 &   3 &   0.08$\pm$0.18 &   0 &        ...      &   1 &   0.20$\pm$0.09\\
1--94        &   0 &        ...      &   0 &        ...      &   3 &   0.29$\pm$0.21 &   1 &   0.24$\pm$0.08 &   0 &        ...     \\
1--60        &   0 &        ...      &   0 &        ...      &   3 &   0.55$\pm$0.08 &   0 &        ...      &   0 &        ...     \\
1--59        &   0 &        ...      &   0 &        ...      &   3 &   0.38$\pm$0.16 &   1 &   0.31$\pm$0.08 &   0 &        ...     \\
G53476\_4543 &   0 &        ...      &   1 & --0.12$\pm$0.08 &   3 &   0.23$\pm$0.16 &   1 &   0.19$\pm$0.08 &   1 &   0.33$\pm$0.09\\
2--160       &   0 &        ...      &   1 &   0.22$\pm$0.08 &   3 &   0.25$\pm$0.17 &   1 &   0.12$\pm$0.08 &   0 &        ...     \\
G53447\_4707 &   1 & --0.23$\pm$0.09 &   0 &        ...      &   3 &   0.26$\pm$0.17 &   1 & --0.02$\pm$0.08 &   1 &   0.25$\pm$0.09\\
G53445\_4647 &   0 &        ...      &   0 &        ...      &   2 &   0.35$\pm$0.20 &   0 &        ...      &   0 &        ...     \\
G53447\_4703 &   0 &        ...      &   0 &        ...      &   3 &   0.15$\pm$0.19 &   0 &        ...      &   0 &        ...     \\
G53425\_4612 &   0 &        ...      &   0 &        ...      &   3 &   0.41$\pm$0.09 &   0 &        ...      &   0 &        ...     \\
G53477\_4539 &   0 &        ...      &   0 &        ...      &   3 &   0.22$\pm$0.16 &   1 &   0.41$\pm$0.08 &   0 &        ...     \\
G53457\_4709 &   0 &        ...      &   0 &        ...      &   3 &   0.22$\pm$0.19 &   0 &        ...      &   0 &        ...     \\
G53391\_4628 &   0 &        ...      &   0 &        ...      &   3 &   0.33$\pm$0.23 &   0 &        ...      &   0 &        ...     \\
G53417\_4431 &   0 &        ...      &   0 &        ...      &   3 &   0.22$\pm$0.13 &   0 &        ...      &   0 &        ...     \\
G53392\_4624 &   0 &        ...      &   0 &        ...      &   2 &   0.61$\pm$0.32 &   0 &        ...      &   0 &        ...     \\
G53414\_4435 &   0 &        ...      &   0 &        ...      &   3 &   0.33$\pm$0.09 &   0 &        ...      &   0 &        ...     
\enddata                                          
\end{deluxetable}                                 

\clearpage 

\begin{deluxetable}{lrrrrr}                       
\tablenum{6}                                      
\tablewidth{0pt}                                  
\tablecaption{Sensitivity of Abundance.\label{tab6}}           
\tablehead{                                       
\colhead{} &                                      
 \colhead{$\Delta$\ew} &                          
 \colhead{$\Delta$\teff} &                        
 \colhead{$\Delta$\grav} &                        
 \colhead{$\Delta$\mtv} &                         
   \colhead{$\Delta$\fe}\\                        
\colhead{} &                                      
\colhead{10\%} &                                  
\colhead{+ 100 K} &                               
\colhead{+ 0.2 dex} &                             
\colhead{+ 0.2 \kms} &                            
\colhead{+ 0.2 dex} }                             
\startdata                                        
Li I :  & & & & &  \\
5000/2.5/1.0 &  0.10\tablenotemark{a} &  0.13 &  0.01 &  0.00 &  0.01 \\
C I  :  & & & & &  \\
4250/1.0/1.4 &  0.10 & -0.28 &  0.09 & -0.02 & -0.03 \\
5000/2.5/1.0 &  0.07 & -0.20 &  0.09 & -0.01 &  0.00 \\
5500/4.0/0.6 &  0.07 & -0.11 &  0.07 &  0.00 &  0.00 \\
O I  :  & & & & &  \\
4250/1.0/1.4 &  0.08 & -0.12 &  0.09 & -0.01 & -0.03 \\
5000/2.5/1.0 &  0.06 & -0.06 &  0.08 & -0.01 & -0.02 \\
Na I :  & & & & &  \\
4250/1.0/1.4 &  0.17 &  0.09 & -0.01 & -0.07 &  0.01 \\
5000/2.5/1.0 &  0.11 &  0.07 & -0.01 & -0.03 &  0.01 \\
5500/4.0/0.6 &  0.10 &  0.06 & -0.01 &  0.00 &  0.00 \\
Mg I :  & & & & &  \\
4250/1.0/1.4 &  0.12 &  0.01 &  0.00 & -0.04 &  0.00 \\
5000/2.5/1.0 &  0.10 &  0.05 & -0.01 & -0.03 &  0.00 \\
5500/4.0/0.6 &  0.16 &  0.06 & -0.03 & -0.02 &  0.00 \\
Al I :  & & & & &  \\
4250/1.0/1.4 &  0.09 &  0.07 &  0.00 & -0.03 &  0.00 \\
5000/2.5/1.0 &  0.06 &  0.06 &  0.00 & -0.01 &  0.01 \\
Si I :  & & & & &  \\
4250/1.0/1.4 &  0.09 & -0.12 &  0.06 & -0.03 & -0.04 \\
5000/2.5/1.0 &  0.10 & -0.07 &  0.05 & -0.03 & -0.04 \\
5500/4.0/0.6 &  0.08 &  0.00 &  0.01 & -0.02 & -0.01 \\
K I  :  & & & & &  \\
4250/1.0/1.4 &  0.26 &  0.14 & -0.01 & -0.16 &  0.00 \\
5000/2.5/1.0 &  0.26 &  0.12 & -0.06 & -0.07 &  0.00 \\
Ca I :  & & & & &  \\
4250/1.0/1.4 &  0.18 &  0.12 & -0.02 & -0.10 &  0.00 \\
5000/2.5/1.0 &  0.13 &  0.09 & -0.03 & -0.05 &  0.00 \\
5500/4.0/0.6 &  0.16 &  0.08 & -0.05 & -0.02 & -0.01 \\
Sc II:  & & & & &  \\
4250/1.0/1.4 &  0.20 & -0.02 &  0.07 & -0.12 & -0.06 \\
5000/2.5/1.0 &  0.12 & -0.01 &  0.08 & -0.05 & -0.06 \\
5500/4.0/0.6 &  0.10 &  0.00 &  0.08 & -0.01 & -0.05 \\
Ti I :  & & & & &  \\
4250/1.0/1.4 &  0.16 &  0.15 &  0.00 & -0.10 &  0.01 \\
5000/2.5/1.0 &  0.08 &  0.12 &  0.00 & -0.03 &  0.01 \\
5500/4.0/0.6 &  0.08 &  0.09 &  0.00 & -0.01 &  0.01 \\
V I  :  & & & & &  \\
4250/1.0/1.4 &  0.22 &  0.18 &  0.01 & -0.12 &  0.00 \\
5000/2.5/1.0 &  0.09 &  0.15 &  0.00 & -0.02 &  0.01 \\
Cr I :  & & & & &  \\
4250/1.0/1.4 &  0.14 &  0.10 &  0.00 & -0.09 &  0.01 \\
5000/2.5/1.0 &  0.08 &  0.08 &  0.00 & -0.03 &  0.01 \\
Mn I :  & & & & &  \\
4250/1.0/1.4 &  0.20 &  0.08 &  0.01 & -0.07 &  0.00 \\
5000/2.5/1.0 &  0.12 &  0.10 &  0.00 & -0.03 &  0.01 \\
Fe I :  & & & & &  \\
4250/1.0/1.4 &  0.15 &  0.03 &  0.02 & -0.09 & -0.02 \\
5000/2.5/1.0 &  0.11 &  0.08 &  0.00 & -0.06 &  0.00 \\
5500/4.0/0.6 &  0.14 &  0.08 & -0.01 & -0.03 &  0.00 \\
Fe II:  & & & & &  \\
4250/1.0/1.4 &  0.09 & -0.20 &  0.13 & -0.04 & -0.09 \\
5000/2.5/1.0 &  0.09 & -0.14 &  0.11 & -0.04 & -0.08 \\
5500/4.0/0.6 &  0.08 & -0.05 &  0.09 & -0.03 & -0.05 \\
Co I :  & & & & &  \\
4250/1.0/1.4 &  0.09 &  0.04 &  0.04 & -0.02 & -0.04 \\
5000/2.5/1.0 &  0.05 &  0.10 &  0.01 &  0.00 & -0.01 \\
Ni I :  & & & & &  \\
4250/1.0/1.4 &  0.20 &  0.01 &  0.04 & -0.10 & -0.04 \\
5000/2.5/1.0 &  0.11 &  0.07 &  0.01 & -0.06 &  0.00 \\
5500/4.0/0.6 &  0.13 &  0.05 &  0.00 & -0.03 & -0.01 \\
Cu I :  & & & & &  \\
4250/1.0/1.4 &  0.26 &  0.05 &  0.05 & -0.10 & -0.03 \\
5000/2.5/1.0 &  0.13 &  0.11 &  0.01 & -0.03 &  0.00 \\
Zn I :  & & & & &  \\
4250/1.0/1.4 &  0.06 & -0.03 &  0.05 & -0.01 & -0.02 \\
Y II :  & & & & &  \\
4250/1.0/1.4 &  0.07 & -0.03 &  0.08 & -0.03 & -0.06 \\
Zr I :  & & & & &  \\
4250/1.0/1.4 &  0.13 &  0.20 &  0.01 & -0.09 &  0.00 \\
5000/2.5/1.0 &  0.05 &  0.15 &  0.00 & -0.01 &  0.01 \\
Ba II:  & & & & &  \\
4250/1.0/1.4 &  0.23 &  0.01 &  0.06 & -0.19 & -0.08 \\
5000/2.5/1.0 &  0.20 &  0.02 &  0.04 & -0.16 & -0.07 \\
5500/4.0/0.6 &  0.21 &  0.03 &  0.03 & -0.10 & -0.07 \\
La II:  & & & & &  \\
4250/1.0/1.4 &  0.08 &  0.01 &  0.08 & -0.04 & -0.07 \\
5000/2.5/1.0 &  0.05 &  0.01 &  0.08 & -0.01 & -0.07 \\
Eu II:  & & & & &  \\
4250/1.0/1.4 &  0.09 & -0.02 &  0.08 & -0.06 & -0.07 \\
5000/2.5/1.0 &  0.06 &  0.00 &  0.08 & -0.02 & -0.06 \\
\enddata                                          
\tablenotetext{a}{Estimated error in the synthesis.}
\end{deluxetable}                                 
 
\clearpage
%
%
\begin{deluxetable}{lccccc}
\tablenum{7}
\tablewidth{0pt}
\tablecaption{Mean Abundance Ratios and Errors 
\label{tab7}}
\tablehead{\colhead{} & \colhead{\# stars} & \colhead{$<$[X/Fe]$>$} 
& \colhead{$\sigma_{obs}$} & \colhead{$\sigma_{pred}$} &
\colhead{$\Delta_{max}$} \\
\colhead{} & \colhead{} & \colhead{(dex)} & \colhead{(dex)} &
\colhead{(dex)} & \colhead{(dex)} }
\startdata
Li\tablenotemark{a} &  3 & +1.10 & 0.17 & 0.13 &   ...           \\
C  &  6 & +1.30\tablenotemark{b} & 0.49\tablenotemark{b} & 0.13 
       & 1.00 $\pm$ 0.37\tablenotemark{b} \\
O  & 25 & +0.19 &  0.18 &  0.12 &  0.48 $\pm$ 0.10 \\
Na & 25 & +0.21 &  0.13 &  0.10 &  0.23 $\pm$ 0.11 \\
Mg & 24 & +0.36 &  0.09 &  0.12 &  0.18 $\pm$ 0.07 \\
Al & 11 & +0.24 &  0.10 &  0.10 &  0.11 $\pm$ 0.10 \\
Si & 24 & +0.28 &  0.14 &  0.17 &  0.19 $\pm$ 0.12 \\
K  & 11 &--0.17 &  0.30 &  0.29 &  0.52 $\pm$ 0.25 \\
Ca & 25 & +0.43 &  0.05 &  0.11 &  0.03 $\pm$ 0.04 \\
Sc & 25 & +0.05 &  0.16 &  0.15 &  0.37 $\pm$ 0.10 \\
Ti & 25 & +0.20 &  0.08 &  0.11 &  0.08 $\pm$ 0.07 \\
V  & 21 & +0.11 &  0.14 &  0.17 &  0.09 $\pm$ 0.08 \\
Cr & 23 &--0.13 &  0.09 &  0.09 &  0.05 $\pm$ 0.05 \\
Mn & 13 &--0.27 &  0.11 &  0.11 &  0.03 $\pm$ 0.08 \\
Co & 17 & +0.04 &  0.05 &  0.10 &  0.06 $\pm$ 0.04 \\
Ni & 25 & +0.01 &  0.06 &  0.09 &  0.16 $\pm$ 0.03 \\
Cu & 21 & +0.07 &  0.14 &  0.21 &  0.19 $\pm$ 0.10 \\
Zn &  8 & +0.46 &  0.16 &  0.22 &  0.29 $\pm$ 0.10 \\
Y  &  3 &--0.22 &  0.04 &  0.15 &  0.06 $\pm$ 0.08 \\
Zr & 10 &--0.14 &  0.25 &  0.16 &  0.61 $\pm$ 0.24 \\
Ba & 25 & +0.34 &  0.12 &  0.19 &  0.03 $\pm$ 0.09 \\
La & 14 & +0.20 &  0.10 &  0.15 &  0.02 $\pm$ 0.10 \\
Eu & 11 & +0.31 &  0.15 &  0.15 &  0.31 $\pm$ 0.11 \\
\enddata
\tablenotetext{a}{For Li, log $\epsilon$(Li) (H=12.0 dex) is given.  For
all other elements, [X/Fe] is given.}
\tablenotetext{b}{The C abundances determined from C I lines 
in the cooler M71 stars are believed
to be spurious due to contamination by lines from the red system of CN.}
\end{deluxetable}

\end{document}